\documentclass[aps,pra,nolongbibliography,notitlepage,twocolumn,superscriptaddress]{revtex4-2}
\usepackage{amsmath}
\usepackage{amssymb}
\usepackage{bbold}
\usepackage{color}
\usepackage{tikz}
\usepackage{appendix}
\usepackage{pgfplots}
\pgfplotsset{compat=1.5}
\usepackage{graphicx}
\usepackage[colorlinks=true, linkcolor=blue, citecolor=blue, pdfencoding=auto]{hyperref}
\usepackage{blindtext}
\usepackage{physics}

\newcommand{\moire}{moir\'e}
\newcommand{\Moire}{Moir\'e}

\def\phdag{{\phantom \dagger}}

\begin{document}
\title{
Local basis for interacting topological bands
}
\author{Nishchhal \surname{Verma} }
\email{nishchhal.verma@columbia.edu}
\affiliation{Department of Physics, Columbia University, New York, NY 10027, USA}

\author{Raquel \surname{Queiroz} }
\affiliation{Department of Physics, Columbia University, New York, NY 10027, USA}
\affiliation{Center for Computational Quantum Physics, Flatiron Institute, New York, New York 10010, USA}

\begin{abstract}
The discovery of correlated states in {\moire} materials has challenged the established methods of projecting interactions into a local Wannier basis due to topological obstructions that manifest in extended interactions. This difficulty can sometimes be evaded by decomposing the band into a basis of extended itinerant states and a lattice of local states, using the heavy fermion prescription.
We revisit this framework by systematically identifying the dominant interaction channels guided by the eigenvalues of the projected density operator. 
This approach can be applied both to tight-binding and continuum models, allowing us to identify a hierarchy in interaction scales that can be universally used to reduce the Hilbert space dimension and determine an appropriate local basis for modeling electronic correlations in interacting topological materials.
\end{abstract}

\maketitle

\paragraph*{Introduction.---}
Correlated phases emerge from the competition between electronic kinetic energy and electron-electron interactions. The kinetic energy favors delocalization, whereas electron repulsion favors spatial localization. 
This wave-particle dichotomy is evident in the Hubbard model \cite{Hubbard1963}, where the kinetic term is diagonal in the plane-wave Bloch basis, interactions are local in real space Wannier basis and their interplay gives rise to a rich landscape of many-body phases \cite{Arovas2022Hubbard, Imada1998MIT}. 
With the two competing terms diagonal in different basis, constructing an effective low-energy model requires projecting the Coulomb interaction onto a subset of Bloch bands. This can be done either perturbatively in terms of interaction matrix elements that scatter Bloch states, or by constructing lattice models using Wannier functions onto which the interactions are projected \cite{Marzari2012MLWF}. 
The latter approach is commonly used in strongly correlated systems where interactions dominate the energetic competition.

Topological obstructions prevent constructing exponentially localized Wannier functions that preserve the local symmetries of the material \cite{BrouderPRL2007, SoluyanovPRB2011, SoluyanovPRB2012, Monaco2018}, challenging this longstanding framework of modeling electronic correlations. This difficulty is particularly evident in {\Moire } materials \cite{CaoNature2018, CaoNature2018a, Andrei2021} which show multiple correlated states when the bands are flattened by the {\moire} potential\cite{Lopes2007, Shallcross2010, bistritzer2011moire, Lopes2012, TBG1}. 
The {\moire} bands at the magic angle of twisted bilayer graphene (TBG) are topological \cite{Zou2018, song2019all, TBG2} and the resulting projected interactions are extended \cite{Koshino2018, Kang2018, TBG3}.
Several works have approached this problem by projecting Coulomb interactions onto Bloch states within the Hartree-Fock approximation \cite{Xie2020, Zhang2020, Bultinck2020, TBG3, TBG4}. Others have explored the map of {\moire} flat bands to Landau levels \cite{tarnopolsky2019origin} in inhomogeneous magnetic fields \cite{ledwith2020fractional, crepel2024topologically, Parhizkar2024, liu2019pseudo}, to utilize Haldane pseudopotential methods \cite{wang2021exact} and outline novel routes to superconductivity \cite{khalaf2021charged}.
While the theoretical approaches have some inconsistencies with experiments \cite{Nuckolls2023Quantum}, including strain appears to improve the agreement \cite{parker2021strain, kwan2021kekule}.

Alongside the Hartree-Fock and Landau level approaches, the ``topological heavy fermion model'' of TBG has provided a key conceptual advancement to understand the correlated states \cite{Kang2021H, Shi2022, song2022magic, Datta2023Heavy}. 
The model was motivated both by evidence of local moments \cite{Xie2019Spectroscopic, Wong2020Cascade} and entropy \cite{rozen2021entropic} in STM, and a symmetry analysis of the wavefunctions at high symmetry points \cite{song2022magic, Calugaru2023}. The heavy fermion model proposes a basis of local $f$ electrons and Dirac itinerant $c$ electrons, which weakly couple, reproducing accurately the wavefunctions of the flat bands at the magic angle. The $f$ electrons experience a strong Hubbard repulsion, while the $c$ electrons interact weakly.
This approach reduces the dimension of the Hilbert space, while showing excellent agreement with Hartree-Fock calculations in the full continuum model.

This brings us to the central question of this work: \emph{what governs the choice of a local basis and the resulting low-energy interactions for bands with topological obstructions?} 
Given the rapid progress in {\moire} materials, which continue to reveal correlated phases in graphene systems \cite{khalaf2019magic, wang2022hierarchy, guerci2022higher, Kang2023, Vafek2023}, transition metal dichalcogenides (TMDs) \cite{Xu2020Correlated, Guo2024Superconductivity} and hybrid heterostructures \cite{park2023observation}, it is important to find a systematic way to choose the appropriate basis to capture interactions in topological bands. With the same goal, there have been other proposals, including coherent states \cite{li2024constraints}, reduced Wannier basis \cite{cole2024reduced}, and supercell Wannier functions \cite{Monsen2024Supercell}. 
In our work, we take a complementary approach based on finding the dominant modes of the projected density operator. Since interactions couple directly to charge density, working in this basis is arguably the most natural choice. For a topological band, we show that there is always more than one relevant mode due to irremovable zeros in the mode wave-function. 
We demonstrate that the modes of the projected density operator yield a local basis that outlines a hierarchical sequence of projected interactions. 
This scheme can be applied equally to tight-binding and continuum models.

The paper is organized as follows. We begin by reviewing the projected density operator and its significance in stabilizing fractionalized phases. Next, we explain the physical significance of the spectrum of this operator in tight-binding models, focusing on the Haldane model. We then discuss continuum models by examining the the multifold fermion model and the flat band in twisted bilayer graphene as examples. Building on this, we introduce a quantitative measure for the number of density modes required to capture the quantum geometry of a band faithfully, determining the complexity of a topological band  -- ranging from two modes such as in the topological phase of the Haldane model to the infinite mode limit of Landau levels.
Finally, we discuss the implications of our approach to the relevant interaction scales for a class of density-density interactions.

\begin{figure}
    \centering
    \includegraphics[width=\linewidth]{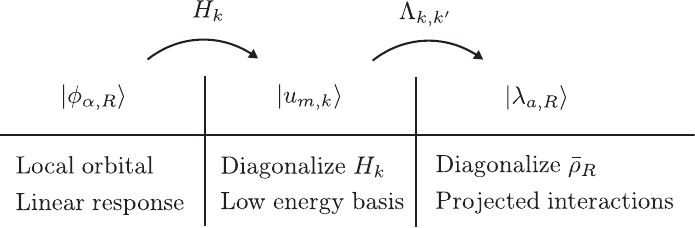}
    \caption{Choice of Basis. The atomic basis $|\phi_{\alpha, R}\rangle$ with microscopic orbitals are the natural choice for describing charge response since vector potential coupling can be efficiently captured via Peierls phase substitution. The Bloch basis $|u_{m,k}\rangle$ diagonalizes the non-interacting Bloch Hamiltonian $H_k$ and is a natural choice for describing low-energy physics in weakly-interacting systems. In this work, we identify a new local basis $|\lambda_{a,R}\rangle$ that diagonalizes the projected density operator and is best suited for representing projected interactions in strongly interacting topological bands.
    }
    \label{fig:schematic}
\end{figure}

\paragraph*{Projected Density Operator.---}
The projected density operator, $\bar{\rho}_q$, is sensitive to the quantum geometry of the projected manifold and takes its most recognizable form in the lowest Landau level (LLL), where these operators follow the Girvin-MacDonald-Platzman (GMP) algebra \cite{Girvin1987}. 
The algebra is an essential feature of the Landau level phenomenology that induces the so-called UV-IR mixing \cite{Haldane2011} and leads to novel many-body states with fractionalized excitations \cite{HalperinJain2020}.

Electrons in crystalline material behave differently from Landau levels.
The starting point is the non-interacting limit where, as a consequence of Bloch's theorem, the total electronic wavefunction is split into a plane-wave and a cell-periodic part with band index $n$ and crystal momenta $k$, $|\psi_{n,k} \rangle = e^{i k\cdot \hat{r} } | u_{n,k} \rangle$.
With this separation of electronic states into bands, the projected density operator for a single band, say $n$, is defined as
\begin{equation}
    \bar{\rho}_q = \sum\limits_{k} \langle u_{n,k} | u_{n,k+q} \rangle c^\dag_{n,k} c^\phdag_{n,k+q}, \label{eq:def:proj_rho}
\end{equation}
where $c^\phdag_{n,k}$ are fermion operators for the target band and the form factors $\langle u_{n,k} | u_{n,k+q} \rangle$ encode the quantum geometry of the single particle states.
Deviations from the LLL can be seen explicitly in this expression. 
While the LLL wavefunctions satisfy $\langle u_k | u_{k+q} \rangle = \exp(-i q\wedge k/2 - q^2/4)$, arbitrary wavefunctions in a lattice do not have this exact form, though they can satisfy it for small $q$ and $q^\prime$ \cite{Roy2014, LEDWITH2021168646}. 
The extent to which the GMP algebra is obeyed depends on the details of the band structure, leading to the definition of ideal band geometry \cite{ledwith2020fractional, wang2021exact}.

The focus of this work lies away from the small-$q$ expansion and the related quantum geometric tensor. We consider the total projected density in the unit cell and its eigen-decomposition
\begin{equation}
    \bar{\rho}_R = \sum\limits_q \bar{\rho}_q = \sum\limits_{k,k^\prime} [\Lambda]_{k,k^\prime} c^\dag_{n,k} c^\phdag_{n,k^\prime} = \sum\limits_a \lambda_a \bar{\rho}_{a,R} \label{eq:proj:rho:svd}
\end{equation}
with eigenvalues $\lambda_a$ and modes $\bar{\rho}_{a,R} \equiv c^\dag_{a,R} c^\phdag_{a,R}$ (see SI for additional details). The form factors are captured by the matrix $\Lambda$ with elements $[\Lambda]_{k,k^\prime} = \langle u_k | u_{k^\prime} \rangle$.
These modes capture the charge density of a subset of states within a band with an envelope function
\begin{equation}
    c^\phdag_{a,R} = \sum\limits_k e^{i k\cdot R} s_{a,k} c^\phdag_{n,k} \label{def:envelope:fnc}
\end{equation}
where $s_{a,k}$ is an eigenvector of $\Lambda$ matrix.
Together, these modes describes the set of orbitals needed to fully capture the charge density of the band.
These modes are conceptually similar to \emph{natural orbitals} in quantum chemistry that diagonalize the one-particle reduced density matrix \cite{davidson1976reduced}. 
However, our formalism is entirely constructed from single-particle states, whereas the quantum chemistry approach includes interactions and many-body effects.

We stress that the formalism includes not only infinitesimal overlaps of Bloch states encoded in the quantum geometric tensor \cite{yu2024quantum}, but finite-$q$ overlaps that provide non-perturbative information about the global geometry of the Bloch states.
This information is reflected in the envelope function.
The procedure is numerically expensive because the overlap matrix $\Lambda$ is not sparse and its dimension depends on the momentum-space mesh, however, it is low rank in many cases.

\paragraph*{Tight-binding models.---}
A tight-binding model with $N_\phi$ orbitals is described by the Bloch Hamiltonian $H_k$, where the crystal momentum $k$ spans the Brillouin zone (BZ) defined by reciprocal lattice vectors ${\bf b}_1$ and ${\bf b}_2$. The eigenvector of the Bloch Hamiltonian is the cell-periodic part of the Bloch wavefunction, $|u_{n,k}\rangle$. In what follows, we focus on one isolated band and omit the index $n$.

A key feature of multi-orbital tight-binding models is orbital embedding, which determines the periodicity of Bloch states. As detailed in the SI, our formalism is independent of embedding but depends on the choice of unit cell. Once the unit cell is fixed, we compute $\Lambda$ in the cell-periodic gauge, where $H_{k+G} = H_k$ and $G$ is a reciprocal lattice vector. 
The rank of $\Lambda$ determines the number of orbitals required to span all band states. Since this number cannot exceed the number of tight-binding orbitals $N_\phi$, the rank $R$ satisfies $R \leq N_\phi$.

We consider the Haldane model as a concrete example to study the projected density operator. 
The model is defined on a honeycomb lattice with inversion breaking mass $M$ and sub-lattice dependent flux $\phi$ threaded with second-neighbor hopping $t_2$. The two bands have a finite Chern number whenever $| M - 3\sqrt{3} t_2 \sin\phi | < 1$. We set $t_2 = 1$, $\phi = \pi/2$ and use $M$ to drive the topological phase transition from topological $M=0$ to trivial $M > 3\sqrt{3}$.

\begin{figure}
    \centering
    \includegraphics[width=\linewidth]{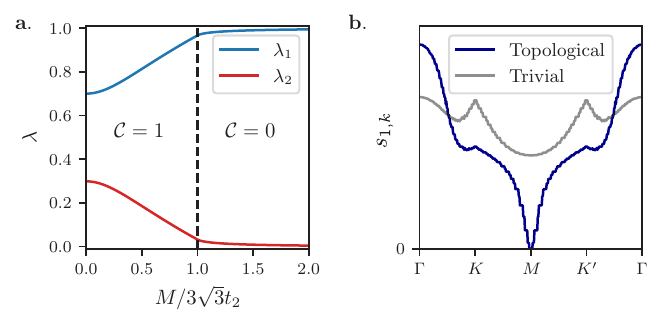}
    \caption{{\bf a}. Eigenvalues of projected density operator in the Haldane model as a function of $M/3\sqrt{3}t_2$ with a topological phase transition indicated with the dashed line. {\bf b}. The envelope function $s_{1,k}$ for the dominant mode along a high symmetry path in the Brillouin zone. The zeros are topologically protected reflecting the impossibility to describe the topological band with a single local mode.
    }
    \label{fig:haldane}
\end{figure}

The behavior of eigenvalues is shown in Fig.~\ref{fig:haldane}{\bf a}.
As $M$ increases, one mode starts to dominate, eventually saturating in the trivial sector. 
We can interpret the $M\rightarrow \infty$ limit where $\lambda_1 \rightarrow 1$ as the atomic limit. Here, the wavefunction of the band can be spanned by one mode across the entire Brillouin zone.
At first glance, the eigenvalues do not reveal the topology of the Chern band as they show a gradual dominance of one mode without any singularity at the topological transition.

The topology instead manifests in the momentum distribution of the mode, defined via the envelope function in Eq.~\eqref{def:envelope:fnc}. 
Since there is a topological obstruction to creating a local orbital from a Chern band \cite{BrouderPRL2007}, the envelope function is forced to have zeroes \cite{li2024constraints}.
These zeros are topologically protected and can be seen in Fig.~\ref{fig:haldane}{\bf b}. 
The robustness of the zeroes can also be seen from the recently developed framework of singular connection \cite{Mera2022}.

\paragraph*{Continuum models.---}
While instructive, the utility of spectrum of projected density in tight-binding models is limited as these models inherently assume a predefined local basis. 
The advantage of our construction lies in continuum models, which are defined in momentum space and lack a clear specification of real-space orbitals. 
In this setting, the spectrum of projected density operator can be used to define a minimal local representation.
This is important because in contrast to tight-binding models where non-zero eigenvalues are upper bounded by the number of orbitals, bands in continuum models do not have such bounds. 
They thus require a more quantitative description of the spectrum.

Here, we introduce two such measures. Since the charge density is normalized such that $\sum_\alpha \lambda_\alpha = 1$, the eigenvalues $\lambda_\alpha$ can be perceived as a probability distribution. To quantify their relative spread, we define the inverse participation ratio $\mathcal{I}$ and the Shannon entropy $\mathcal{S}$
\begin{equation}
    \mathcal{I} = \sum\limits_{\alpha=1}^N \lambda_\alpha^2, \quad  \mathcal{S} = -\dfrac{1}{ \log N }\sum\limits_{\alpha=1}^N \lambda_\alpha \log \lambda_\alpha. \label{eq:def-I-S}
\end{equation}
The inverse participation ratio $\mathcal{I}$ quantifies localization, with $\mathcal{I} \to 1$ for a single dominant mode and $\mathcal{I} \to 0$ when many modes contribute. The Shannon entropy $\mathcal{S}$ provides a complementary measure, approaching $0$ for a sharply localized distribution and $1$ for a maximally spread one. 
A band with scalar plane waves wavefunctions has a single mode with $\mathcal{I}=1, \mathcal{S}=0$, while Landau levels where infinitely many modes contribute equally have $\mathcal{I}=0, \mathcal{S}=1$. As we will see in the following, bands in {\moire} materials fall between the two extremes, but quantitatively, they are well captured by a small number of modes.

\begin{figure}
    \centering
    \includegraphics[width=\linewidth]{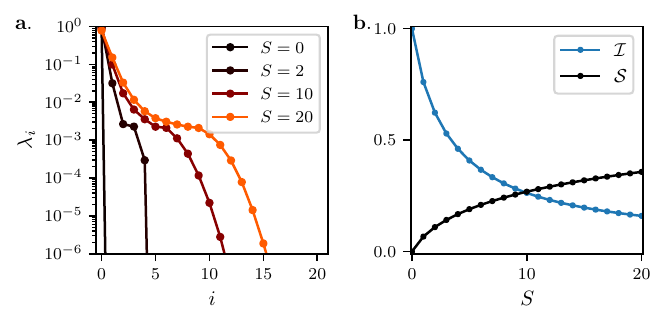}
    \caption{Spectrum of the projected density operator in the multifold fermion continuum model.
    The two measures, $\mathcal{I}$ and $\mathcal{S}$ are defined in Eq.~\eqref{eq:def-I-S}. The IPR $\mathcal{I}$ quantifies the dominance of one mode and entropy $\mathcal{S}$ quantifies the delocalization across multiple modes. The two asymptotically approach $\mathcal{I}=0$ and $\mathcal{S}=1$ in the large $S$ limit which corresponds to an ideal band.}
    \label{fig:multifold}
\end{figure}

We now turn to a concrete model to show how these two measures connect to the quantum geometry of the band. Motivated by the multifold fermion \cite{wieder2016double, bradlyn2016beyond}, we consider the continuum model
\begin{equation}
    H_k = d_k \cdot \sigma, \quad d_k = \left( k_x, k_y, \dfrac{k^2}{2M} + 1 \right)
\end{equation}
where $\sigma$ represents the $2S+1$ dimensional representation of the $SU(2)$ group. For a given $S$, the model has $2S+1$ bands with an energetically isolated top band with Chern number $\mathcal{C} = 2S$. 
The model thus hosts an ``infinite Chern'' band in the large $S$ limit \cite{Tan2024}. Given the simplicity of the model, the overlaps can be computed analytically
\begin{equation}
    \langle u_k | u_{k^\prime} \rangle = \dfrac{ (M^2 + k\cdot k^\prime - i k \wedge k^\prime)^{2S} }{ (M^2 + k^2)^{S} (M^2 + (k^\prime)^2)^{S} }
\end{equation}
and depend on parameters $S$ and $M$ which capture two different properties.
The spin $S$ determines the number of band inversions, reflecting how many microscopic orbitals hybridize to form the band. It is therefore reflected in the Chern number. On the other hand, $M$ sets the momentum-space scale over which the wavefunctions vary. It does not change the Chern number but controls the spread of the Berry curvature.
In addition to these parameters, the cutoff plays a crucial role as it determines the proportion of the momentum space where the effects of band inversion are reflected in the wavefunction. Unlike tight-binding models with a natural momentum cutoff, continuum models require an explicit choice. 
Here, we set the cutoff $\Lambda_k \gg M$, fix $M$ and vary $S$.

As shown in Fig.~\ref{fig:multifold}{\bf a}, the number of non-zero eigenvalues increases with $S$, reflecting the notion that more orbitals are needed to describe bands formed via higher number of band inversions. The model interpolates between a single-orbital limit ($S=0$) and the ideal band regime ($S \rightarrow \infty$)~\cite{Tan2024}, with an intermediate phase characterized by a finite set of contributing modes. This is where the measures in Eq.~\eqref{eq:def-I-S} become informative. As shown in Fig.~\ref{fig:multifold}{\bf b}, the participation ratio decreases with $S$, indicating multiple relevant modes, but it does not capture the distribution among them. That nuance is revealed by the entropy, which increases when the spectral weight is shared evenly.

\paragraph*{Projected Interactions.---}
Identifying the relevant interaction in the low-energy manifold is highly non-trivial.
Even local density-density interactions are known to project down to non-local four-fermion interactions depending on the spread of the Wannier function of the low-energy band \cite{torma2022}.
In what follows, we restrict the discussion to the UV interaction of the type $H_{\rm int} = \sum_{R,R^\prime} V_{R-R^\prime} \rho_R \rho_{R^\prime}$ where $\rho_R$ is the total density inside unit cell $R$ including all internal degrees of freedom.
Upon projection, the interaction can take various forms, but we limit ourselves to the density-density channel, with projected interaction $\bar{H}_{\rm int} = \sum_{R,R^\prime} V_{R-R^\prime} \bar{\rho}_R \bar{\rho}_{R^\prime}$.

Once these approximations are in place, we can employ the decomposition in Eq.~\eqref{eq:proj:rho:svd} to write the projected interactions as
\begin{align}
    \bar{H}_{\rm int} = \sum\limits_{R,R^\prime,\alpha\beta}V_{R-R^\prime} \lambda_\alpha\lambda_\beta c^\dag_{R,\alpha} c^\phdag_{R,\alpha} c^\dag_{R^\prime,\beta} c^\phdag_{R^\prime,\beta}
\end{align}
in terms of the modes $\alpha$ and $\beta$. This formalism makes the interplay between topology and correlations explicit. Topology determines the minimum number of modes, while the strength of these values dictates the interaction scale. 

A trivial band can be described by a single mode, whereas a Landau level requires multiple modes. 
This is the reason why projected interactions have non-trivial phenomenology in Landau levels but not in trivial flat bands.
Extending this perspective to the heavy fermion model \cite{song2022magic}, we expect two modes: a dominant $\lambda_D$ associated with the $f$ electrons and a sub-dominant $\lambda_S$ associated with the $c$ electrons. The relevant interaction scales then follow $U_f = V \lambda_D^2 > U_{cf} = V \lambda_D \lambda_S > U_c = V \lambda_S^2$ where $U_f$ is the interaction between $f$ electrons, $U_c$ is the interaction between $c$ electrons, and $U_{cf}$ is the inter-orbital interaction.  
This framework suggests that the heavy fermion picture can be universal, requiring only the existence of a single dominant eigenvalue \cite{Yu2023, Herzog-Arbeitman2024}.

\begin{figure}
    \centering
    \includegraphics[width=\linewidth]{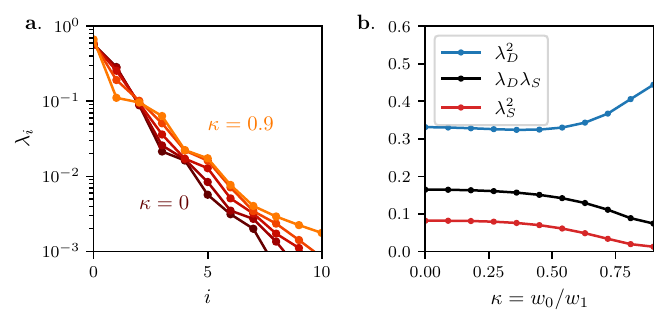}
    \caption{{\bf a}.~Spectrum of projected density in TBG flat-band close to magic angle for different values of the chiral parameter $\kappa = w_0/w_1$. {\bf b}. Leading projected density-density interactions between the dominant and sub-dominant modes.}
    \label{fig:TBG}
\end{figure}

We now apply this scheme to twisted bilayer graphene (TBG). The spin- and valley-resolved band structure is captured by the Bistritzer-MacDonald model \cite{bistritzer2011moire}, and features two effective Dirac fields near the $K, K'$ points of the {\moire} Brillouin zone, each rotated by twist angle $\pm {\theta}/{2}$ and coupled via an inter-layer {\moire } potential
\begin{align}
H = \begin{pmatrix}
    -iv \sigma_{\theta/2} \cdot \nabla & T(r) \\ 
    T^{\dag}(r) & -iv \sigma_{-\theta/2} \cdot \nabla  
\end{pmatrix}, \label{contmod}
\end{align}
where $\sigma_{\theta/2} = e^{-i\theta/4 \sigma_{z}} (\sigma_x, \sigma_y) e^{i\theta/4 \sigma_{z}}$, $\nabla = (\partial_x, \partial_y)$, and $T(r) =  \sum_{j=1}^{3} T_{j} e^{-i q_j \cdot r}$ where $q$ are determined by the {\moire } potential \cite{tarnopolsky2019origin} and
\begin{align}
T_{j+1} = w_0 \sigma_0 + w_1 \big( \cos(2\pi j/3) \sigma_x + \sin(2\pi j/3) \sigma_y \big), \label{Tj}
\end{align} 
where $w_0$ controls $AA$ tunneling, while $w_1$ governs $AB$ and $BA$ tunnelings.
We fix the twist angle $\theta = 1.15^\circ$, introduce a small sublattice mass to split the flat bands and focus on the lower band.
The spectrum of the projected density operator is shown in Fig.~\ref{fig:TBG}{\bf a}.~for different values of the chiral parameter $\kappa = w_0/w_1$. There is a clear dominant mode $\lambda_D$ irrespective of $\kappa$, leading to a dominant interaction channel $\lambda_D^2$, consistent with the heavy fermion representation of TBG. 
This approximation gets better as $\kappa$ approaches to 1 leading to a hierarchy of interaction channels with both dominant and sub-dominant interactions shown in Fig.~\ref{fig:TBG}{\bf b}.

Note that while topology ensures at least two relevant modes it does not constrain their relative strengths. 
This is where quantum geometry, or specifically the distribution of Berry curvature, becomes important and can tune the relative strength of the dominant and subdominant interaction channels without altering the topological invariants of the system.
This was recently demonstrated in the context of TBG with a model of Dirac fermions in an inhomogeneous magnetic field \cite{ledwith2024nonlocal}.

\paragraph*{Discussion.---}
The spectrum of the projected density operator provides a route to organize electronic interactions in topological bands. 
This scheme unifies several complementary approaches in {\moire} materials, from Hubbard and Landau level to heavy fermion models.
In our work, we first employed tight-binding models to establish the connection between the spectrum of the density operator and topology. 
Using the Haldane model, we showed that while topology dictates the presence of at least two nonzero eigenvalues, the momentum distribution includes topologically protected zeros.
Next, we extended this analysis to continuum models, using multifold fermions as a tunable platform to outline quantitative measures for the spectrum.
Finally, applying our formalism to TBG flat band, we found one dominant mode in agreement with the heavy fermion model.
In the context of {\moire} materials, our key finding is that the projected density operator generically exhibits low rank. This is perhaps expected because previous studies on collective excitations have found low-rank scattering matrices~\cite{TBG5}, which are similar but not identical to our formalism.

In broad terms, our work highlights the significance of global quantum geometry in correlated systems, as opposed to the quantum geometric tensor. 
It also opens up several promising directions for future investigations: generalizing the formalism to other interaction channels beyond density-density; extending it to multiple bands where the topological obstruction is tied to symmetry; and comparing this basis with the Wannier basis in strongly correlated systems other than {\moire} materials.

\paragraph*{Acknowledgement.---}
This work is supported as part of Programmable Quantum Materials, an Energy Frontier Research Center funded by the U.S. Department of Energy (DOE), Office of Science, Basic Energy Sciences (BES), under award DE-SC0019443, and the NSF CAREER award No. DMR-2340394.
The Flatiron Institute is a division of the Simons Foundation.

\bibliography{minBasis}

\begin{thebibliography}{71}%
\makeatletter
\providecommand \@ifxundefined [1]{%
 \@ifx{#1\undefined}
}%
\providecommand \@ifnum [1]{%
 \ifnum #1\expandafter \@firstoftwo
 \else \expandafter \@secondoftwo
 \fi
}%
\providecommand \@ifx [1]{%
 \ifx #1\expandafter \@firstoftwo
 \else \expandafter \@secondoftwo
 \fi
}%
\providecommand \natexlab [1]{#1}%
\providecommand \enquote  [1]{``#1''}%
\providecommand \bibnamefont  [1]{#1}%
\providecommand \bibfnamefont [1]{#1}%
\providecommand \citenamefont [1]{#1}%
\providecommand \href@noop [0]{\@secondoftwo}%
\providecommand \href [0]{\begingroup \@sanitize@url \@href}%
\providecommand \@href[1]{\@@startlink{#1}\@@href}%
\providecommand \@@href[1]{\endgroup#1\@@endlink}%
\providecommand \@sanitize@url [0]{\catcode `\\12\catcode `\$12\catcode `\&12\catcode `\#12\catcode `\^12\catcode `\_12\catcode `\%12\relax}%
\providecommand \@@startlink[1]{}%
\providecommand \@@endlink[0]{}%
\providecommand \url  [0]{\begingroup\@sanitize@url \@url }%
\providecommand \@url [1]{\endgroup\@href {#1}{\urlprefix }}%
\providecommand \urlprefix  [0]{URL }%
\providecommand \Eprint [0]{\href }%
\providecommand \doibase [0]{https://doi.org/}%
\providecommand \selectlanguage [0]{\@gobble}%
\providecommand \bibinfo  [0]{\@secondoftwo}%
\providecommand \bibfield  [0]{\@secondoftwo}%
\providecommand \translation [1]{[#1]}%
\providecommand \BibitemOpen [0]{}%
\providecommand \bibitemStop [0]{}%
\providecommand \bibitemNoStop [0]{.\EOS\space}%
\providecommand \EOS [0]{\spacefactor3000\relax}%
\providecommand \BibitemShut  [1]{\csname bibitem#1\endcsname}%
\let\auto@bib@innerbib\@empty
\bibitem [{\citenamefont {Hubbard}(1963)}]{Hubbard1963}%
  \BibitemOpen
  \bibfield  {author} {\bibinfo {author} {\bibfnamefont {J.}~\bibnamefont {Hubbard}},\ }\href {https://doi.org/10.1098/rspa.1963.0204} {\bibfield  {journal} {\bibinfo  {journal} {Proc. R. Soc. Lond. A Math. Phys. Sci.}\ }\textbf {\bibinfo {volume} {276}},\ \bibinfo {pages} {238} (\bibinfo {year} {1963})}\BibitemShut {NoStop}%
\bibitem [{\citenamefont {Arovas}\ \emph {et~al.}(2022)\citenamefont {Arovas}, \citenamefont {Berg}, \citenamefont {Kivelson},\ and\ \citenamefont {Raghu}}]{Arovas2022Hubbard}%
  \BibitemOpen
  \bibfield  {author} {\bibinfo {author} {\bibfnamefont {D.~P.}\ \bibnamefont {Arovas}}, \bibinfo {author} {\bibfnamefont {E.}~\bibnamefont {Berg}}, \bibinfo {author} {\bibfnamefont {S.~A.}\ \bibnamefont {Kivelson}},\ and\ \bibinfo {author} {\bibfnamefont {S.}~\bibnamefont {Raghu}},\ }\href {https://doi.org/10.1146/annurev-conmatphys-031620-102024} {\bibfield  {journal} {\bibinfo  {journal} {Annu. Rev. Condens. Matter Phys.}\ }\textbf {\bibinfo {volume} {13}},\ \bibinfo {pages} {239} (\bibinfo {year} {2022})}\BibitemShut {NoStop}%
\bibitem [{\citenamefont {Imada}\ \emph {et~al.}(1998)\citenamefont {Imada}, \citenamefont {Fujimori},\ and\ \citenamefont {Tokura}}]{Imada1998MIT}%
  \BibitemOpen
  \bibfield  {author} {\bibinfo {author} {\bibfnamefont {M.}~\bibnamefont {Imada}}, \bibinfo {author} {\bibfnamefont {A.}~\bibnamefont {Fujimori}},\ and\ \bibinfo {author} {\bibfnamefont {Y.}~\bibnamefont {Tokura}},\ }\href {https://doi.org/10.1103/RevModPhys.70.1039} {\bibfield  {journal} {\bibinfo  {journal} {Rev. Mod. Phys.}\ }\textbf {\bibinfo {volume} {70}},\ \bibinfo {pages} {1039} (\bibinfo {year} {1998})}\BibitemShut {NoStop}%
\bibitem [{\citenamefont {Marzari}\ \emph {et~al.}(2012)\citenamefont {Marzari}, \citenamefont {Mostofi}, \citenamefont {Yates}, \citenamefont {Souza},\ and\ \citenamefont {Vanderbilt}}]{Marzari2012MLWF}%
  \BibitemOpen
  \bibfield  {author} {\bibinfo {author} {\bibfnamefont {N.}~\bibnamefont {Marzari}}, \bibinfo {author} {\bibfnamefont {A.~A.}\ \bibnamefont {Mostofi}}, \bibinfo {author} {\bibfnamefont {J.~R.}\ \bibnamefont {Yates}}, \bibinfo {author} {\bibfnamefont {I.}~\bibnamefont {Souza}},\ and\ \bibinfo {author} {\bibfnamefont {D.}~\bibnamefont {Vanderbilt}},\ }\href {https://doi.org/10.1103/RevModPhys.84.1419} {\bibfield  {journal} {\bibinfo  {journal} {Rev. Mod. Phys.}\ }\textbf {\bibinfo {volume} {84}},\ \bibinfo {pages} {1419} (\bibinfo {year} {2012})}\BibitemShut {NoStop}%
\bibitem [{\citenamefont {Brouder}\ \emph {et~al.}(2007)\citenamefont {Brouder}, \citenamefont {Panati}, \citenamefont {Calandra}, \citenamefont {Mourougane},\ and\ \citenamefont {Marzari}}]{BrouderPRL2007}%
  \BibitemOpen
  \bibfield  {author} {\bibinfo {author} {\bibfnamefont {C.}~\bibnamefont {Brouder}}, \bibinfo {author} {\bibfnamefont {G.}~\bibnamefont {Panati}}, \bibinfo {author} {\bibfnamefont {M.}~\bibnamefont {Calandra}}, \bibinfo {author} {\bibfnamefont {C.}~\bibnamefont {Mourougane}},\ and\ \bibinfo {author} {\bibfnamefont {N.}~\bibnamefont {Marzari}},\ }\href {https://doi.org/10.1103/PhysRevLett.98.046402} {\bibfield  {journal} {\bibinfo  {journal} {Phys. Rev. Lett.}\ }\textbf {\bibinfo {volume} {98}},\ \bibinfo {pages} {046402} (\bibinfo {year} {2007})}\BibitemShut {NoStop}%
\bibitem [{\citenamefont {Soluyanov}\ and\ \citenamefont {Vanderbilt}(2011)}]{SoluyanovPRB2011}%
  \BibitemOpen
  \bibfield  {author} {\bibinfo {author} {\bibfnamefont {A.~A.}\ \bibnamefont {Soluyanov}}\ and\ \bibinfo {author} {\bibfnamefont {D.}~\bibnamefont {Vanderbilt}},\ }\href {https://doi.org/10.1103/PhysRevB.83.035108} {\bibfield  {journal} {\bibinfo  {journal} {Phys. Rev. B}\ }\textbf {\bibinfo {volume} {83}},\ \bibinfo {pages} {035108} (\bibinfo {year} {2011})}\BibitemShut {NoStop}%
\bibitem [{\citenamefont {Soluyanov}\ and\ \citenamefont {Vanderbilt}(2012)}]{SoluyanovPRB2012}%
  \BibitemOpen
  \bibfield  {author} {\bibinfo {author} {\bibfnamefont {A.~A.}\ \bibnamefont {Soluyanov}}\ and\ \bibinfo {author} {\bibfnamefont {D.}~\bibnamefont {Vanderbilt}},\ }\href {https://doi.org/10.1103/PhysRevB.85.115415} {\bibfield  {journal} {\bibinfo  {journal} {Phys. Rev. B}\ }\textbf {\bibinfo {volume} {85}},\ \bibinfo {pages} {115415} (\bibinfo {year} {2012})}\BibitemShut {NoStop}%
\bibitem [{\citenamefont {Monaco}\ \emph {et~al.}(2018)\citenamefont {Monaco}, \citenamefont {Panati}, \citenamefont {Pisante},\ and\ \citenamefont {Teufel}}]{Monaco2018}%
  \BibitemOpen
  \bibfield  {author} {\bibinfo {author} {\bibfnamefont {D.}~\bibnamefont {Monaco}}, \bibinfo {author} {\bibfnamefont {G.}~\bibnamefont {Panati}}, \bibinfo {author} {\bibfnamefont {A.}~\bibnamefont {Pisante}},\ and\ \bibinfo {author} {\bibfnamefont {S.}~\bibnamefont {Teufel}},\ }\href {https://doi.org/10.1007/s00220-017-3067-7} {\bibfield  {journal} {\bibinfo  {journal} {Communications in Mathematical Physics}\ }\textbf {\bibinfo {volume} {359}},\ \bibinfo {pages} {61} (\bibinfo {year} {2018})}\BibitemShut {NoStop}%
\bibitem [{\citenamefont {Cao}\ \emph {et~al.}(2018{\natexlab{a}})\citenamefont {Cao}, \citenamefont {Fatemi}, \citenamefont {Fang}, \citenamefont {Watanabe}, \citenamefont {Taniguchi}, \citenamefont {Kaxiras},\ and\ \citenamefont {{Jarillo-Herrero}}}]{CaoNature2018}%
  \BibitemOpen
  \bibfield  {author} {\bibinfo {author} {\bibfnamefont {Y.}~\bibnamefont {Cao}}, \bibinfo {author} {\bibfnamefont {V.}~\bibnamefont {Fatemi}}, \bibinfo {author} {\bibfnamefont {S.}~\bibnamefont {Fang}}, \bibinfo {author} {\bibfnamefont {K.}~\bibnamefont {Watanabe}}, \bibinfo {author} {\bibfnamefont {T.}~\bibnamefont {Taniguchi}}, \bibinfo {author} {\bibfnamefont {E.}~\bibnamefont {Kaxiras}},\ and\ \bibinfo {author} {\bibfnamefont {P.}~\bibnamefont {{Jarillo-Herrero}}},\ }\href {https://doi.org/10.1038/nature26160} {\bibfield  {journal} {\bibinfo  {journal} {Nature}\ }\textbf {\bibinfo {volume} {556}},\ \bibinfo {pages} {43} (\bibinfo {year} {2018}{\natexlab{a}})}\BibitemShut {NoStop}%
\bibitem [{\citenamefont {Cao}\ \emph {et~al.}(2018{\natexlab{b}})\citenamefont {Cao}, \citenamefont {Fatemi}, \citenamefont {Demir}, \citenamefont {Fang}, \citenamefont {Tomarken}, \citenamefont {Luo}, \citenamefont {{Sanchez-Yamagishi}}, \citenamefont {Watanabe}, \citenamefont {Taniguchi}, \citenamefont {Kaxiras}, \citenamefont {Ashoori},\ and\ \citenamefont {{Jarillo-Herrero}}}]{CaoNature2018a}%
  \BibitemOpen
  \bibfield  {author} {\bibinfo {author} {\bibfnamefont {Y.}~\bibnamefont {Cao}}, \bibinfo {author} {\bibfnamefont {V.}~\bibnamefont {Fatemi}}, \bibinfo {author} {\bibfnamefont {A.}~\bibnamefont {Demir}}, \bibinfo {author} {\bibfnamefont {S.}~\bibnamefont {Fang}}, \bibinfo {author} {\bibfnamefont {S.~L.}\ \bibnamefont {Tomarken}}, \bibinfo {author} {\bibfnamefont {J.~Y.}\ \bibnamefont {Luo}}, \bibinfo {author} {\bibfnamefont {J.~D.}\ \bibnamefont {{Sanchez-Yamagishi}}}, \bibinfo {author} {\bibfnamefont {K.}~\bibnamefont {Watanabe}}, \bibinfo {author} {\bibfnamefont {T.}~\bibnamefont {Taniguchi}}, \bibinfo {author} {\bibfnamefont {E.}~\bibnamefont {Kaxiras}}, \bibinfo {author} {\bibfnamefont {R.~C.}\ \bibnamefont {Ashoori}},\ and\ \bibinfo {author} {\bibfnamefont {P.}~\bibnamefont {{Jarillo-Herrero}}},\ }\href {https://doi.org/10.1038/nature26154} {\bibfield  {journal} {\bibinfo  {journal} {Nature}\ }\textbf {\bibinfo {volume} {556}},\ \bibinfo {pages} {80} (\bibinfo {year} {2018}{\natexlab{b}})}\BibitemShut
  {NoStop}%
\bibitem [{\citenamefont {Andrei}\ \emph {et~al.}(2021)\citenamefont {Andrei}, \citenamefont {Efetov}, \citenamefont {Jarillo-Herrero}, \citenamefont {MacDonald}, \citenamefont {Mak}, \citenamefont {Senthil}, \citenamefont {Tutuc}, \citenamefont {Yazdani},\ and\ \citenamefont {Young}}]{Andrei2021}%
  \BibitemOpen
  \bibfield  {author} {\bibinfo {author} {\bibfnamefont {E.~Y.}\ \bibnamefont {Andrei}}, \bibinfo {author} {\bibfnamefont {D.~K.}\ \bibnamefont {Efetov}}, \bibinfo {author} {\bibfnamefont {P.}~\bibnamefont {Jarillo-Herrero}}, \bibinfo {author} {\bibfnamefont {A.~H.}\ \bibnamefont {MacDonald}}, \bibinfo {author} {\bibfnamefont {K.~F.}\ \bibnamefont {Mak}}, \bibinfo {author} {\bibfnamefont {T.}~\bibnamefont {Senthil}}, \bibinfo {author} {\bibfnamefont {E.}~\bibnamefont {Tutuc}}, \bibinfo {author} {\bibfnamefont {A.}~\bibnamefont {Yazdani}},\ and\ \bibinfo {author} {\bibfnamefont {A.~F.}\ \bibnamefont {Young}},\ }\href {https://doi.org/10.1038/s41578-021-00284-1} {\bibfield  {journal} {\bibinfo  {journal} {Nature Reviews Materials}\ }\textbf {\bibinfo {volume} {6}},\ \bibinfo {pages} {201} (\bibinfo {year} {2021})}\BibitemShut {NoStop}%
\bibitem [{\citenamefont {Lopes~dos Santos}\ \emph {et~al.}(2007)\citenamefont {Lopes~dos Santos}, \citenamefont {Peres},\ and\ \citenamefont {Castro~Neto}}]{Lopes2007}%
  \BibitemOpen
  \bibfield  {author} {\bibinfo {author} {\bibfnamefont {J.~M.~B.}\ \bibnamefont {Lopes~dos Santos}}, \bibinfo {author} {\bibfnamefont {N.~M.~R.}\ \bibnamefont {Peres}},\ and\ \bibinfo {author} {\bibfnamefont {A.~H.}\ \bibnamefont {Castro~Neto}},\ }\href {https://doi.org/10.1103/PhysRevLett.99.256802} {\bibfield  {journal} {\bibinfo  {journal} {Phys. Rev. Lett.}\ }\textbf {\bibinfo {volume} {99}},\ \bibinfo {pages} {256802} (\bibinfo {year} {2007})}\BibitemShut {NoStop}%
\bibitem [{\citenamefont {Shallcross}\ \emph {et~al.}(2010)\citenamefont {Shallcross}, \citenamefont {Sharma}, \citenamefont {Kandelaki},\ and\ \citenamefont {Pankratov}}]{Shallcross2010}%
  \BibitemOpen
  \bibfield  {author} {\bibinfo {author} {\bibfnamefont {S.}~\bibnamefont {Shallcross}}, \bibinfo {author} {\bibfnamefont {S.}~\bibnamefont {Sharma}}, \bibinfo {author} {\bibfnamefont {E.}~\bibnamefont {Kandelaki}},\ and\ \bibinfo {author} {\bibfnamefont {O.~A.}\ \bibnamefont {Pankratov}},\ }\href {https://doi.org/10.1103/PhysRevB.81.165105} {\bibfield  {journal} {\bibinfo  {journal} {Phys. Rev. B}\ }\textbf {\bibinfo {volume} {81}},\ \bibinfo {pages} {165105} (\bibinfo {year} {2010})}\BibitemShut {NoStop}%
\bibitem [{\citenamefont {Bistritzer}\ and\ \citenamefont {MacDonald}(2011)}]{bistritzer2011moire}%
  \BibitemOpen
  \bibfield  {author} {\bibinfo {author} {\bibfnamefont {R.}~\bibnamefont {Bistritzer}}\ and\ \bibinfo {author} {\bibfnamefont {A.~H.}\ \bibnamefont {MacDonald}},\ }\href {https://www.pnas.org/doi/full/10.1073/pnas.1108174108} {\bibfield  {journal} {\bibinfo  {journal} {Proceedings of the National Academy of Sciences}\ }\textbf {\bibinfo {volume} {108}},\ \bibinfo {pages} {12233} (\bibinfo {year} {2011})}\BibitemShut {NoStop}%
\bibitem [{\citenamefont {Lopes~dos Santos}\ \emph {et~al.}(2012)\citenamefont {Lopes~dos Santos}, \citenamefont {Peres},\ and\ \citenamefont {Castro~Neto}}]{Lopes2012}%
  \BibitemOpen
  \bibfield  {author} {\bibinfo {author} {\bibfnamefont {J.~M.~B.}\ \bibnamefont {Lopes~dos Santos}}, \bibinfo {author} {\bibfnamefont {N.~M.~R.}\ \bibnamefont {Peres}},\ and\ \bibinfo {author} {\bibfnamefont {A.~H.}\ \bibnamefont {Castro~Neto}},\ }\href {https://doi.org/10.1103/PhysRevB.86.155449} {\bibfield  {journal} {\bibinfo  {journal} {Phys. Rev. B}\ }\textbf {\bibinfo {volume} {86}},\ \bibinfo {pages} {155449} (\bibinfo {year} {2012})}\BibitemShut {NoStop}%
\bibitem [{\citenamefont {Bernevig}\ \emph {et~al.}(2021{\natexlab{a}})\citenamefont {Bernevig}, \citenamefont {Song}, \citenamefont {Regnault}, \citenamefont {Lian} \emph {et~al.}}]{TBG1}%
  \BibitemOpen
  \bibfield  {author} {\bibinfo {author} {\bibfnamefont {B.~A.}\ \bibnamefont {Bernevig}}, \bibinfo {author} {\bibfnamefont {Z.-D.}\ \bibnamefont {Song}}, \bibinfo {author} {\bibfnamefont {N.}~\bibnamefont {Regnault}}, \bibinfo {author} {\bibfnamefont {B.}~\bibnamefont {Lian}}, \emph {et~al.},\ }\href {https://journals.aps.org/prb/abstract/10.1103/PhysRevB.103.205411} {\bibfield  {journal} {\bibinfo  {journal} {Phys. Rev. B}\ }\textbf {\bibinfo {volume} {103}},\ \bibinfo {pages} {205411} (\bibinfo {year} {2021}{\natexlab{a}})}\BibitemShut {NoStop}%
\bibitem [{\citenamefont {Zou}\ \emph {et~al.}(2018)\citenamefont {Zou}, \citenamefont {Po}, \citenamefont {Vishwanath},\ and\ \citenamefont {Senthil}}]{Zou2018}%
  \BibitemOpen
  \bibfield  {author} {\bibinfo {author} {\bibfnamefont {L.}~\bibnamefont {Zou}}, \bibinfo {author} {\bibfnamefont {H.~C.}\ \bibnamefont {Po}}, \bibinfo {author} {\bibfnamefont {A.}~\bibnamefont {Vishwanath}},\ and\ \bibinfo {author} {\bibfnamefont {T.}~\bibnamefont {Senthil}},\ }\href {https://doi.org/10.1103/PhysRevB.98.085435} {\bibfield  {journal} {\bibinfo  {journal} {Phys. Rev. B}\ }\textbf {\bibinfo {volume} {98}},\ \bibinfo {pages} {085435} (\bibinfo {year} {2018})}\BibitemShut {NoStop}%
\bibitem [{\citenamefont {Song}\ \emph {et~al.}(2019)\citenamefont {Song}, \citenamefont {Wang}, \citenamefont {Shi}, \citenamefont {Li}, \citenamefont {Fang},\ and\ \citenamefont {Bernevig}}]{song2019all}%
  \BibitemOpen
  \bibfield  {author} {\bibinfo {author} {\bibfnamefont {Z.}~\bibnamefont {Song}}, \bibinfo {author} {\bibfnamefont {Z.}~\bibnamefont {Wang}}, \bibinfo {author} {\bibfnamefont {W.}~\bibnamefont {Shi}}, \bibinfo {author} {\bibfnamefont {G.}~\bibnamefont {Li}}, \bibinfo {author} {\bibfnamefont {C.}~\bibnamefont {Fang}},\ and\ \bibinfo {author} {\bibfnamefont {B.~A.}\ \bibnamefont {Bernevig}},\ }\href {https://journals.aps.org/prl/abstract/10.1103/PhysRevLett.123.036401} {\bibfield  {journal} {\bibinfo  {journal} {Phys. Rev. Lett.}\ }\textbf {\bibinfo {volume} {123}},\ \bibinfo {pages} {036401} (\bibinfo {year} {2019})}\BibitemShut {NoStop}%
\bibitem [{\citenamefont {Bernevig}\ \emph {et~al.}(2021{\natexlab{b}})\citenamefont {Bernevig}, \citenamefont {Song}, \citenamefont {Regnault}, \citenamefont {Lian} \emph {et~al.}}]{TBG2}%
  \BibitemOpen
  \bibfield  {author} {\bibinfo {author} {\bibfnamefont {B.~A.}\ \bibnamefont {Bernevig}}, \bibinfo {author} {\bibfnamefont {Z.-D.}\ \bibnamefont {Song}}, \bibinfo {author} {\bibfnamefont {N.}~\bibnamefont {Regnault}}, \bibinfo {author} {\bibfnamefont {B.}~\bibnamefont {Lian}}, \emph {et~al.},\ }\href {https://journals.aps.org/prb/abstract/10.1103/PhysRevB.103.205412} {\bibfield  {journal} {\bibinfo  {journal} {Phys. Rev. B}\ }\textbf {\bibinfo {volume} {103}},\ \bibinfo {pages} {205412} (\bibinfo {year} {2021}{\natexlab{b}})}\BibitemShut {NoStop}%
\bibitem [{\citenamefont {Koshino}\ \emph {et~al.}(2018)\citenamefont {Koshino}, \citenamefont {Yuan}, \citenamefont {Koretsune}, \citenamefont {Ochi}, \citenamefont {Kuroki},\ and\ \citenamefont {Fu}}]{Koshino2018}%
  \BibitemOpen
  \bibfield  {author} {\bibinfo {author} {\bibfnamefont {M.}~\bibnamefont {Koshino}}, \bibinfo {author} {\bibfnamefont {N.~F.~Q.}\ \bibnamefont {Yuan}}, \bibinfo {author} {\bibfnamefont {T.}~\bibnamefont {Koretsune}}, \bibinfo {author} {\bibfnamefont {M.}~\bibnamefont {Ochi}}, \bibinfo {author} {\bibfnamefont {K.}~\bibnamefont {Kuroki}},\ and\ \bibinfo {author} {\bibfnamefont {L.}~\bibnamefont {Fu}},\ }\href {https://doi.org/10.1103/PhysRevX.8.031087} {\bibfield  {journal} {\bibinfo  {journal} {Phys. Rev. X}\ }\textbf {\bibinfo {volume} {8}},\ \bibinfo {pages} {031087} (\bibinfo {year} {2018})}\BibitemShut {NoStop}%
\bibitem [{\citenamefont {Kang}\ and\ \citenamefont {Vafek}(2018)}]{Kang2018}%
  \BibitemOpen
  \bibfield  {author} {\bibinfo {author} {\bibfnamefont {J.}~\bibnamefont {Kang}}\ and\ \bibinfo {author} {\bibfnamefont {O.}~\bibnamefont {Vafek}},\ }\href {https://doi.org/10.1103/PhysRevX.8.031088} {\bibfield  {journal} {\bibinfo  {journal} {Phys. Rev. X}\ }\textbf {\bibinfo {volume} {8}},\ \bibinfo {pages} {031088} (\bibinfo {year} {2018})}\BibitemShut {NoStop}%
\bibitem [{\citenamefont {Bernevig}\ \emph {et~al.}(2021{\natexlab{c}})\citenamefont {Bernevig}, \citenamefont {Song}, \citenamefont {Regnault}, \citenamefont {Lian} \emph {et~al.}}]{TBG3}%
  \BibitemOpen
  \bibfield  {author} {\bibinfo {author} {\bibfnamefont {B.~A.}\ \bibnamefont {Bernevig}}, \bibinfo {author} {\bibfnamefont {Z.-D.}\ \bibnamefont {Song}}, \bibinfo {author} {\bibfnamefont {N.}~\bibnamefont {Regnault}}, \bibinfo {author} {\bibfnamefont {B.}~\bibnamefont {Lian}}, \emph {et~al.},\ }\href {https://journals.aps.org/prb/abstract/10.1103/PhysRevB.103.205413} {\bibfield  {journal} {\bibinfo  {journal} {Phys. Rev. B}\ }\textbf {\bibinfo {volume} {103}},\ \bibinfo {pages} {205413} (\bibinfo {year} {2021}{\natexlab{c}})}\BibitemShut {NoStop}%
\bibitem [{\citenamefont {Xie}\ and\ \citenamefont {MacDonald}(2020)}]{Xie2020}%
  \BibitemOpen
  \bibfield  {author} {\bibinfo {author} {\bibfnamefont {M.}~\bibnamefont {Xie}}\ and\ \bibinfo {author} {\bibfnamefont {A.~H.}\ \bibnamefont {MacDonald}},\ }\href {https://doi.org/10.1103/PhysRevLett.124.097601} {\bibfield  {journal} {\bibinfo  {journal} {Phys. Rev. Lett.}\ }\textbf {\bibinfo {volume} {124}},\ \bibinfo {pages} {097601} (\bibinfo {year} {2020})}\BibitemShut {NoStop}%
\bibitem [{\citenamefont {Zhang}\ \emph {et~al.}(2020)\citenamefont {Zhang}, \citenamefont {Jiang}, \citenamefont {Wang},\ and\ \citenamefont {Zhang}}]{Zhang2020}%
  \BibitemOpen
  \bibfield  {author} {\bibinfo {author} {\bibfnamefont {Y.}~\bibnamefont {Zhang}}, \bibinfo {author} {\bibfnamefont {K.}~\bibnamefont {Jiang}}, \bibinfo {author} {\bibfnamefont {Z.}~\bibnamefont {Wang}},\ and\ \bibinfo {author} {\bibfnamefont {F.}~\bibnamefont {Zhang}},\ }\href {https://doi.org/10.1103/PhysRevB.102.035136} {\bibfield  {journal} {\bibinfo  {journal} {Phys. Rev. B}\ }\textbf {\bibinfo {volume} {102}},\ \bibinfo {pages} {035136} (\bibinfo {year} {2020})}\BibitemShut {NoStop}%
\bibitem [{\citenamefont {Bultinck}\ \emph {et~al.}(2020)\citenamefont {Bultinck}, \citenamefont {Khalaf}, \citenamefont {Liu}, \citenamefont {Chatterjee}, \citenamefont {Vishwanath},\ and\ \citenamefont {Zaletel}}]{Bultinck2020}%
  \BibitemOpen
  \bibfield  {author} {\bibinfo {author} {\bibfnamefont {N.}~\bibnamefont {Bultinck}}, \bibinfo {author} {\bibfnamefont {E.}~\bibnamefont {Khalaf}}, \bibinfo {author} {\bibfnamefont {S.}~\bibnamefont {Liu}}, \bibinfo {author} {\bibfnamefont {S.}~\bibnamefont {Chatterjee}}, \bibinfo {author} {\bibfnamefont {A.}~\bibnamefont {Vishwanath}},\ and\ \bibinfo {author} {\bibfnamefont {M.~P.}\ \bibnamefont {Zaletel}},\ }\href {https://doi.org/10.1103/PhysRevX.10.031034} {\bibfield  {journal} {\bibinfo  {journal} {Phys. Rev. X}\ }\textbf {\bibinfo {volume} {10}},\ \bibinfo {pages} {031034} (\bibinfo {year} {2020})}\BibitemShut {NoStop}%
\bibitem [{\citenamefont {Bernevig}\ \emph {et~al.}(2021{\natexlab{d}})\citenamefont {Bernevig}, \citenamefont {Song}, \citenamefont {Regnault}, \citenamefont {Lian} \emph {et~al.}}]{TBG4}%
  \BibitemOpen
  \bibfield  {author} {\bibinfo {author} {\bibfnamefont {B.~A.}\ \bibnamefont {Bernevig}}, \bibinfo {author} {\bibfnamefont {Z.-D.}\ \bibnamefont {Song}}, \bibinfo {author} {\bibfnamefont {N.}~\bibnamefont {Regnault}}, \bibinfo {author} {\bibfnamefont {B.}~\bibnamefont {Lian}}, \emph {et~al.},\ }\href {https://journals.aps.org/prb/abstract/10.1103/PhysRevB.103.205414} {\bibfield  {journal} {\bibinfo  {journal} {Phys. Rev. B}\ }\textbf {\bibinfo {volume} {103}},\ \bibinfo {pages} {205414} (\bibinfo {year} {2021}{\natexlab{d}})}\BibitemShut {NoStop}%
\bibitem [{\citenamefont {Tarnopolsky}\ \emph {et~al.}(2019)\citenamefont {Tarnopolsky}, \citenamefont {Kruchkov},\ and\ \citenamefont {Vishwanath}}]{tarnopolsky2019origin}%
  \BibitemOpen
  \bibfield  {author} {\bibinfo {author} {\bibfnamefont {G.}~\bibnamefont {Tarnopolsky}}, \bibinfo {author} {\bibfnamefont {A.~J.}\ \bibnamefont {Kruchkov}},\ and\ \bibinfo {author} {\bibfnamefont {A.}~\bibnamefont {Vishwanath}},\ }\href@noop {} {\bibfield  {journal} {\bibinfo  {journal} {Phys. Rev. Lett.}\ }\textbf {\bibinfo {volume} {122}},\ \bibinfo {pages} {106405} (\bibinfo {year} {2019})}\BibitemShut {NoStop}%
\bibitem [{\citenamefont {Ledwith}\ \emph {et~al.}(2020)\citenamefont {Ledwith}, \citenamefont {Tarnopolsky}, \citenamefont {Khalaf},\ and\ \citenamefont {Vishwanath}}]{ledwith2020fractional}%
  \BibitemOpen
  \bibfield  {author} {\bibinfo {author} {\bibfnamefont {P.~J.}\ \bibnamefont {Ledwith}}, \bibinfo {author} {\bibfnamefont {G.}~\bibnamefont {Tarnopolsky}}, \bibinfo {author} {\bibfnamefont {E.}~\bibnamefont {Khalaf}},\ and\ \bibinfo {author} {\bibfnamefont {A.}~\bibnamefont {Vishwanath}},\ }\href@noop {} {\bibfield  {journal} {\bibinfo  {journal} {Phys. Rev. Res.}\ }\textbf {\bibinfo {volume} {2}},\ \bibinfo {pages} {023237} (\bibinfo {year} {2020})}\BibitemShut {NoStop}%
\bibitem [{\citenamefont {Cr{\'e}pel}\ \emph {et~al.}(2024)\citenamefont {Cr{\'e}pel}, \citenamefont {Ding}, \citenamefont {Verma}, \citenamefont {Regnault},\ and\ \citenamefont {Queiroz}}]{crepel2024topologically}%
  \BibitemOpen
  \bibfield  {author} {\bibinfo {author} {\bibfnamefont {V.}~\bibnamefont {Cr{\'e}pel}}, \bibinfo {author} {\bibfnamefont {P.}~\bibnamefont {Ding}}, \bibinfo {author} {\bibfnamefont {N.}~\bibnamefont {Verma}}, \bibinfo {author} {\bibfnamefont {N.}~\bibnamefont {Regnault}},\ and\ \bibinfo {author} {\bibfnamefont {R.}~\bibnamefont {Queiroz}},\ }\href {https://arxiv.org/abs/2403.19656} {\bibfield  {journal} {\bibinfo  {journal} {arXiv preprint arXiv:2403.19656}\ } (\bibinfo {year} {2024})}\BibitemShut {NoStop}%
\bibitem [{\citenamefont {Parhizkar}\ and\ \citenamefont {Galitski}(2024)}]{Parhizkar2024}%
  \BibitemOpen
  \bibfield  {author} {\bibinfo {author} {\bibfnamefont {A.}~\bibnamefont {Parhizkar}}\ and\ \bibinfo {author} {\bibfnamefont {V.}~\bibnamefont {Galitski}},\ }\href {https://doi.org/10.1103/PhysRevB.110.L121111} {\bibfield  {journal} {\bibinfo  {journal} {Phys. Rev. B}\ }\textbf {\bibinfo {volume} {110}},\ \bibinfo {pages} {L121111} (\bibinfo {year} {2024})}\BibitemShut {NoStop}%
\bibitem [{\citenamefont {Liu}\ \emph {et~al.}(2019)\citenamefont {Liu}, \citenamefont {Liu},\ and\ \citenamefont {Dai}}]{liu2019pseudo}%
  \BibitemOpen
  \bibfield  {author} {\bibinfo {author} {\bibfnamefont {J.}~\bibnamefont {Liu}}, \bibinfo {author} {\bibfnamefont {J.}~\bibnamefont {Liu}},\ and\ \bibinfo {author} {\bibfnamefont {X.}~\bibnamefont {Dai}},\ }\href {https://doi.org/10.1103/PhysRevB.99.155415} {\bibfield  {journal} {\bibinfo  {journal} {Phys. Rev. B}\ }\textbf {\bibinfo {volume} {99}},\ \bibinfo {pages} {155415} (\bibinfo {year} {2019})}\BibitemShut {NoStop}%
\bibitem [{\citenamefont {Wang}\ \emph {et~al.}(2021)\citenamefont {Wang}, \citenamefont {Cano}, \citenamefont {Millis}, \citenamefont {Liu},\ and\ \citenamefont {Yang}}]{wang2021exact}%
  \BibitemOpen
  \bibfield  {author} {\bibinfo {author} {\bibfnamefont {J.}~\bibnamefont {Wang}}, \bibinfo {author} {\bibfnamefont {J.}~\bibnamefont {Cano}}, \bibinfo {author} {\bibfnamefont {A.~J.}\ \bibnamefont {Millis}}, \bibinfo {author} {\bibfnamefont {Z.}~\bibnamefont {Liu}},\ and\ \bibinfo {author} {\bibfnamefont {B.}~\bibnamefont {Yang}},\ }\href {https://doi.org/10.1103/PhysRevLett.127.246403} {\bibfield  {journal} {\bibinfo  {journal} {Phys. Rev. Lett.}\ }\textbf {\bibinfo {volume} {127}},\ \bibinfo {pages} {246403} (\bibinfo {year} {2021})}\BibitemShut {NoStop}%
\bibitem [{\citenamefont {Khalaf}\ \emph {et~al.}(2021)\citenamefont {Khalaf}, \citenamefont {Chatterjee}, \citenamefont {Bultinck}, \citenamefont {Zaletel},\ and\ \citenamefont {Vishwanath}}]{khalaf2021charged}%
  \BibitemOpen
  \bibfield  {author} {\bibinfo {author} {\bibfnamefont {E.}~\bibnamefont {Khalaf}}, \bibinfo {author} {\bibfnamefont {S.}~\bibnamefont {Chatterjee}}, \bibinfo {author} {\bibfnamefont {N.}~\bibnamefont {Bultinck}}, \bibinfo {author} {\bibfnamefont {M.~P.}\ \bibnamefont {Zaletel}},\ and\ \bibinfo {author} {\bibfnamefont {A.}~\bibnamefont {Vishwanath}},\ }\href {https://www.science.org/doi/abs/10.1126/sciadv.abf5299} {\bibfield  {journal} {\bibinfo  {journal} {Science advances}\ }\textbf {\bibinfo {volume} {7}},\ \bibinfo {pages} {eabf5299} (\bibinfo {year} {2021})}\BibitemShut {NoStop}%
\bibitem [{\citenamefont {Nuckolls}\ \emph {et~al.}(2023)\citenamefont {Nuckolls}, \citenamefont {Lee}, \citenamefont {Oh}, \citenamefont {Wong}, \citenamefont {Soejima}, \citenamefont {Hong}, \citenamefont {Călugăru}, \citenamefont {Herzog-Arbeitman}, \citenamefont {Bernevig}, \citenamefont {Watanabe}, \citenamefont {Taniguchi}, \citenamefont {Regnault}, \citenamefont {Zaletel},\ and\ \citenamefont {Yazdani}}]{Nuckolls2023Quantum}%
  \BibitemOpen
  \bibfield  {author} {\bibinfo {author} {\bibfnamefont {K.~P.}\ \bibnamefont {Nuckolls}}, \bibinfo {author} {\bibfnamefont {R.~L.}\ \bibnamefont {Lee}}, \bibinfo {author} {\bibfnamefont {M.}~\bibnamefont {Oh}}, \bibinfo {author} {\bibfnamefont {D.}~\bibnamefont {Wong}}, \bibinfo {author} {\bibfnamefont {T.}~\bibnamefont {Soejima}}, \bibinfo {author} {\bibfnamefont {J.~P.}\ \bibnamefont {Hong}}, \bibinfo {author} {\bibfnamefont {D.}~\bibnamefont {Călugăru}}, \bibinfo {author} {\bibfnamefont {J.}~\bibnamefont {Herzog-Arbeitman}}, \bibinfo {author} {\bibfnamefont {B.~A.}\ \bibnamefont {Bernevig}}, \bibinfo {author} {\bibfnamefont {K.}~\bibnamefont {Watanabe}}, \bibinfo {author} {\bibfnamefont {T.}~\bibnamefont {Taniguchi}}, \bibinfo {author} {\bibfnamefont {N.}~\bibnamefont {Regnault}}, \bibinfo {author} {\bibfnamefont {M.~P.}\ \bibnamefont {Zaletel}},\ and\ \bibinfo {author} {\bibfnamefont {A.}~\bibnamefont {Yazdani}},\ }\href {https://doi.org/10.1038/s41586-023-06226-x} {\bibfield  {journal} {\bibinfo
  {journal} {Nature}\ }\textbf {\bibinfo {volume} {620}},\ \bibinfo {pages} {525} (\bibinfo {year} {2023})}\BibitemShut {NoStop}%
\bibitem [{\citenamefont {Parker}\ \emph {et~al.}(2021)\citenamefont {Parker}, \citenamefont {Soejima}, \citenamefont {Hauschild}, \citenamefont {Zaletel}, \citenamefont {Bultinck} \emph {et~al.}}]{parker2021strain}%
  \BibitemOpen
  \bibfield  {author} {\bibinfo {author} {\bibfnamefont {D.~E.}\ \bibnamefont {Parker}}, \bibinfo {author} {\bibfnamefont {T.}~\bibnamefont {Soejima}}, \bibinfo {author} {\bibfnamefont {J.}~\bibnamefont {Hauschild}}, \bibinfo {author} {\bibfnamefont {M.~P.}\ \bibnamefont {Zaletel}}, \bibinfo {author} {\bibfnamefont {N.}~\bibnamefont {Bultinck}}, \emph {et~al.},\ }\href {https://journals.aps.org/prl/abstract/10.1103/PhysRevLett.127.027601} {\bibfield  {journal} {\bibinfo  {journal} {Phys. Rev. Lett.}\ }\textbf {\bibinfo {volume} {127}},\ \bibinfo {pages} {027601} (\bibinfo {year} {2021})}\BibitemShut {NoStop}%
\bibitem [{\citenamefont {Kwan}\ \emph {et~al.}(2021)\citenamefont {Kwan}, \citenamefont {Wagner}, \citenamefont {Soejima}, \citenamefont {Zaletel}, \citenamefont {Simon}, \citenamefont {Parameswaran},\ and\ \citenamefont {Bultinck}}]{kwan2021kekule}%
  \BibitemOpen
  \bibfield  {author} {\bibinfo {author} {\bibfnamefont {Y.~H.}\ \bibnamefont {Kwan}}, \bibinfo {author} {\bibfnamefont {G.}~\bibnamefont {Wagner}}, \bibinfo {author} {\bibfnamefont {T.}~\bibnamefont {Soejima}}, \bibinfo {author} {\bibfnamefont {M.~P.}\ \bibnamefont {Zaletel}}, \bibinfo {author} {\bibfnamefont {S.~H.}\ \bibnamefont {Simon}}, \bibinfo {author} {\bibfnamefont {S.~A.}\ \bibnamefont {Parameswaran}},\ and\ \bibinfo {author} {\bibfnamefont {N.}~\bibnamefont {Bultinck}},\ }\href {https://journals.aps.org/prx/abstract/10.1103/PhysRevX.11.041063} {\bibfield  {journal} {\bibinfo  {journal} {Phys. Rev. X}\ }\textbf {\bibinfo {volume} {11}},\ \bibinfo {pages} {041063} (\bibinfo {year} {2021})}\BibitemShut {NoStop}%
\bibitem [{\citenamefont {Kang}\ \emph {et~al.}(2021)\citenamefont {Kang}, \citenamefont {Bernevig},\ and\ \citenamefont {Vafek}}]{Kang2021H}%
  \BibitemOpen
  \bibfield  {author} {\bibinfo {author} {\bibfnamefont {J.}~\bibnamefont {Kang}}, \bibinfo {author} {\bibfnamefont {B.~A.}\ \bibnamefont {Bernevig}},\ and\ \bibinfo {author} {\bibfnamefont {O.}~\bibnamefont {Vafek}},\ }\href {https://doi.org/10.1103/PhysRevLett.127.266402} {\bibfield  {journal} {\bibinfo  {journal} {Phys. Rev. Lett.}\ }\textbf {\bibinfo {volume} {127}},\ \bibinfo {pages} {266402} (\bibinfo {year} {2021})}\BibitemShut {NoStop}%
\bibitem [{\citenamefont {Shi}\ and\ \citenamefont {Dai}(2022)}]{Shi2022}%
  \BibitemOpen
  \bibfield  {author} {\bibinfo {author} {\bibfnamefont {H.}~\bibnamefont {Shi}}\ and\ \bibinfo {author} {\bibfnamefont {X.}~\bibnamefont {Dai}},\ }\href {https://doi.org/10.1103/PhysRevB.106.245129} {\bibfield  {journal} {\bibinfo  {journal} {Phys. Rev. B}\ }\textbf {\bibinfo {volume} {106}},\ \bibinfo {pages} {245129} (\bibinfo {year} {2022})}\BibitemShut {NoStop}%
\bibitem [{\citenamefont {Song}\ and\ \citenamefont {Bernevig}(2022)}]{song2022magic}%
  \BibitemOpen
  \bibfield  {author} {\bibinfo {author} {\bibfnamefont {Z.-D.}\ \bibnamefont {Song}}\ and\ \bibinfo {author} {\bibfnamefont {B.~A.}\ \bibnamefont {Bernevig}},\ }\href {https://journals.aps.org/prl/abstract/10.1103/PhysRevLett.129.047601} {\bibfield  {journal} {\bibinfo  {journal} {Phys. Rev. Lett.}\ }\textbf {\bibinfo {volume} {129}},\ \bibinfo {pages} {047601} (\bibinfo {year} {2022})}\BibitemShut {NoStop}%
\bibitem [{\citenamefont {Datta}\ \emph {et~al.}(2023)\citenamefont {Datta}, \citenamefont {Calderón}, \citenamefont {Camjayi},\ and\ \citenamefont {Bascones}}]{Datta2023Heavy}%
  \BibitemOpen
  \bibfield  {author} {\bibinfo {author} {\bibfnamefont {A.}~\bibnamefont {Datta}}, \bibinfo {author} {\bibfnamefont {M.~J.}\ \bibnamefont {Calderón}}, \bibinfo {author} {\bibfnamefont {A.}~\bibnamefont {Camjayi}},\ and\ \bibinfo {author} {\bibfnamefont {E.}~\bibnamefont {Bascones}},\ }\href {https://doi.org/10.1038/s41467-023-40754-4} {\bibfield  {journal} {\bibinfo  {journal} {Nat. Commun.}\ }\textbf {\bibinfo {volume} {14}},\ \bibinfo {pages} {5036} (\bibinfo {year} {2023})}\BibitemShut {NoStop}%
\bibitem [{\citenamefont {Xie}\ \emph {et~al.}(2019)\citenamefont {Xie}, \citenamefont {Lian}, \citenamefont {Jäck}, \citenamefont {Liu}, \citenamefont {Chiu}, \citenamefont {Watanabe}, \citenamefont {Taniguchi}, \citenamefont {Bernevig},\ and\ \citenamefont {Yazdani}}]{Xie2019Spectroscopic}%
  \BibitemOpen
  \bibfield  {author} {\bibinfo {author} {\bibfnamefont {Y.}~\bibnamefont {Xie}}, \bibinfo {author} {\bibfnamefont {B.}~\bibnamefont {Lian}}, \bibinfo {author} {\bibfnamefont {B.}~\bibnamefont {Jäck}}, \bibinfo {author} {\bibfnamefont {X.}~\bibnamefont {Liu}}, \bibinfo {author} {\bibfnamefont {C.-K.}\ \bibnamefont {Chiu}}, \bibinfo {author} {\bibfnamefont {K.}~\bibnamefont {Watanabe}}, \bibinfo {author} {\bibfnamefont {T.}~\bibnamefont {Taniguchi}}, \bibinfo {author} {\bibfnamefont {B.~A.}\ \bibnamefont {Bernevig}},\ and\ \bibinfo {author} {\bibfnamefont {A.}~\bibnamefont {Yazdani}},\ }\href {https://doi.org/10.1038/s41586-019-1422-x} {\bibfield  {journal} {\bibinfo  {journal} {Nature}\ }\textbf {\bibinfo {volume} {572}},\ \bibinfo {pages} {101} (\bibinfo {year} {2019})}\BibitemShut {NoStop}%
\bibitem [{\citenamefont {Wong}\ \emph {et~al.}(2020)\citenamefont {Wong}, \citenamefont {Nuckolls}, \citenamefont {Oh}, \citenamefont {Lian}, \citenamefont {Xie}, \citenamefont {Jeon}, \citenamefont {Watanabe}, \citenamefont {Taniguchi}, \citenamefont {Bernevig},\ and\ \citenamefont {Yazdani}}]{Wong2020Cascade}%
  \BibitemOpen
  \bibfield  {author} {\bibinfo {author} {\bibfnamefont {D.}~\bibnamefont {Wong}}, \bibinfo {author} {\bibfnamefont {K.~P.}\ \bibnamefont {Nuckolls}}, \bibinfo {author} {\bibfnamefont {M.}~\bibnamefont {Oh}}, \bibinfo {author} {\bibfnamefont {B.}~\bibnamefont {Lian}}, \bibinfo {author} {\bibfnamefont {Y.}~\bibnamefont {Xie}}, \bibinfo {author} {\bibfnamefont {S.}~\bibnamefont {Jeon}}, \bibinfo {author} {\bibfnamefont {K.}~\bibnamefont {Watanabe}}, \bibinfo {author} {\bibfnamefont {T.}~\bibnamefont {Taniguchi}}, \bibinfo {author} {\bibfnamefont {B.~A.}\ \bibnamefont {Bernevig}},\ and\ \bibinfo {author} {\bibfnamefont {A.}~\bibnamefont {Yazdani}},\ }\href {https://doi.org/10.1038/s41586-020-2339-0} {\bibfield  {journal} {\bibinfo  {journal} {Nature}\ }\textbf {\bibinfo {volume} {582}},\ \bibinfo {pages} {198} (\bibinfo {year} {2020})}\BibitemShut {NoStop}%
\bibitem [{\citenamefont {Rozen}\ \emph {et~al.}(2021)\citenamefont {Rozen}, \citenamefont {Park}, \citenamefont {Zondiner}, \citenamefont {Cao}, \citenamefont {Rodan-Legrain}, \citenamefont {Taniguchi}, \citenamefont {Watanabe}, \citenamefont {Oreg}, \citenamefont {Stern}, \citenamefont {Berg}, \citenamefont {Jarillo-Herrero},\ and\ \citenamefont {Ilani}}]{rozen2021entropic}%
  \BibitemOpen
  \bibfield  {author} {\bibinfo {author} {\bibfnamefont {A.}~\bibnamefont {Rozen}}, \bibinfo {author} {\bibfnamefont {J.~M.}\ \bibnamefont {Park}}, \bibinfo {author} {\bibfnamefont {U.}~\bibnamefont {Zondiner}}, \bibinfo {author} {\bibfnamefont {Y.}~\bibnamefont {Cao}}, \bibinfo {author} {\bibfnamefont {D.}~\bibnamefont {Rodan-Legrain}}, \bibinfo {author} {\bibfnamefont {T.}~\bibnamefont {Taniguchi}}, \bibinfo {author} {\bibfnamefont {K.}~\bibnamefont {Watanabe}}, \bibinfo {author} {\bibfnamefont {Y.}~\bibnamefont {Oreg}}, \bibinfo {author} {\bibfnamefont {A.}~\bibnamefont {Stern}}, \bibinfo {author} {\bibfnamefont {E.}~\bibnamefont {Berg}}, \bibinfo {author} {\bibfnamefont {P.}~\bibnamefont {Jarillo-Herrero}},\ and\ \bibinfo {author} {\bibfnamefont {S.}~\bibnamefont {Ilani}},\ }\href {https://doi.org/10.1038/s41586-021-03319-3} {\bibfield  {journal} {\bibinfo  {journal} {Nature}\ }\textbf {\bibinfo {volume} {592}},\ \bibinfo {pages} {214} (\bibinfo {year} {2021})}\BibitemShut {NoStop}%
\bibitem [{\citenamefont {C\u{a}lug\u{a}ru}\ \emph {et~al.}(2023)\citenamefont {C\u{a}lug\u{a}ru}, \citenamefont {Borovkov}, \citenamefont {Lau}, \citenamefont {Coleman}, \citenamefont {Song},\ and\ \citenamefont {Bernevig}}]{Calugaru2023}%
  \BibitemOpen
  \bibfield  {author} {\bibinfo {author} {\bibfnamefont {D.}~\bibnamefont {C\u{a}lug\u{a}ru}}, \bibinfo {author} {\bibfnamefont {M.}~\bibnamefont {Borovkov}}, \bibinfo {author} {\bibfnamefont {L.~L.}\ \bibnamefont {Lau}}, \bibinfo {author} {\bibfnamefont {P.}~\bibnamefont {Coleman}}, \bibinfo {author} {\bibfnamefont {Z.-D.}\ \bibnamefont {Song}},\ and\ \bibinfo {author} {\bibfnamefont {B.~A.}\ \bibnamefont {Bernevig}},\ }\href {https://doi.org/10.1063/10.0017625} {\bibfield  {journal} {\bibinfo  {journal} {Low Temperature Physics}\ }\textbf {\bibinfo {volume} {49}},\ \bibinfo {pages} {640} (\bibinfo {year} {2023})}\BibitemShut {NoStop}%
\bibitem [{\citenamefont {Khalaf}\ \emph {et~al.}(2019)\citenamefont {Khalaf}, \citenamefont {Kruchkov}, \citenamefont {Tarnopolsky},\ and\ \citenamefont {Vishwanath}}]{khalaf2019magic}%
  \BibitemOpen
  \bibfield  {author} {\bibinfo {author} {\bibfnamefont {E.}~\bibnamefont {Khalaf}}, \bibinfo {author} {\bibfnamefont {A.~J.}\ \bibnamefont {Kruchkov}}, \bibinfo {author} {\bibfnamefont {G.}~\bibnamefont {Tarnopolsky}},\ and\ \bibinfo {author} {\bibfnamefont {A.}~\bibnamefont {Vishwanath}},\ }\href {https://journals.aps.org/prb/abstract/10.1103/PhysRevB.100.085109} {\bibfield  {journal} {\bibinfo  {journal} {Phys. Rev. B}\ }\textbf {\bibinfo {volume} {100}},\ \bibinfo {pages} {085109} (\bibinfo {year} {2019})}\BibitemShut {NoStop}%
\bibitem [{\citenamefont {Wang}\ and\ \citenamefont {Liu}(2022)}]{wang2022hierarchy}%
  \BibitemOpen
  \bibfield  {author} {\bibinfo {author} {\bibfnamefont {J.}~\bibnamefont {Wang}}\ and\ \bibinfo {author} {\bibfnamefont {Z.}~\bibnamefont {Liu}},\ }\href {https://doi.org/10.1103/PhysRevLett.128.176403} {\bibfield  {journal} {\bibinfo  {journal} {Phys. Rev. Lett.}\ }\textbf {\bibinfo {volume} {128}},\ \bibinfo {pages} {176403} (\bibinfo {year} {2022})}\BibitemShut {NoStop}%
\bibitem [{\citenamefont {Guerci}\ \emph {et~al.}(2022)\citenamefont {Guerci}, \citenamefont {Simon},\ and\ \citenamefont {Mora}}]{guerci2022higher}%
  \BibitemOpen
  \bibfield  {author} {\bibinfo {author} {\bibfnamefont {D.}~\bibnamefont {Guerci}}, \bibinfo {author} {\bibfnamefont {P.}~\bibnamefont {Simon}},\ and\ \bibinfo {author} {\bibfnamefont {C.}~\bibnamefont {Mora}},\ }\href {https://journals.aps.org/prresearch/abstract/10.1103/PhysRevResearch.4.L012013} {\bibfield  {journal} {\bibinfo  {journal} {Phys. Rev. Res.}\ }\textbf {\bibinfo {volume} {4}},\ \bibinfo {pages} {L012013} (\bibinfo {year} {2022})}\BibitemShut {NoStop}%
\bibitem [{\citenamefont {Kang}\ and\ \citenamefont {Vafek}(2023)}]{Kang2023}%
  \BibitemOpen
  \bibfield  {author} {\bibinfo {author} {\bibfnamefont {J.}~\bibnamefont {Kang}}\ and\ \bibinfo {author} {\bibfnamefont {O.}~\bibnamefont {Vafek}},\ }\href {https://doi.org/10.1103/PhysRevB.107.075408} {\bibfield  {journal} {\bibinfo  {journal} {Phys. Rev. B}\ }\textbf {\bibinfo {volume} {107}},\ \bibinfo {pages} {075408} (\bibinfo {year} {2023})}\BibitemShut {NoStop}%
\bibitem [{\citenamefont {Vafek}\ and\ \citenamefont {Kang}(2023)}]{Vafek2023}%
  \BibitemOpen
  \bibfield  {author} {\bibinfo {author} {\bibfnamefont {O.}~\bibnamefont {Vafek}}\ and\ \bibinfo {author} {\bibfnamefont {J.}~\bibnamefont {Kang}},\ }\href {https://doi.org/10.1103/PhysRevB.107.075123} {\bibfield  {journal} {\bibinfo  {journal} {Phys. Rev. B}\ }\textbf {\bibinfo {volume} {107}},\ \bibinfo {pages} {075123} (\bibinfo {year} {2023})}\BibitemShut {NoStop}%
\bibitem [{\citenamefont {Xu}\ \emph {et~al.}(2020)\citenamefont {Xu}, \citenamefont {Wang}, \citenamefont {Liu}, \citenamefont {Watanabe}, \citenamefont {Taniguchi}, \citenamefont {Hone}, \citenamefont {Shan},\ and\ \citenamefont {Mak}}]{Xu2020Correlated}%
  \BibitemOpen
  \bibfield  {author} {\bibinfo {author} {\bibfnamefont {X.}~\bibnamefont {Xu}}, \bibinfo {author} {\bibfnamefont {Z.}~\bibnamefont {Wang}}, \bibinfo {author} {\bibfnamefont {J.}~\bibnamefont {Liu}}, \bibinfo {author} {\bibfnamefont {K.}~\bibnamefont {Watanabe}}, \bibinfo {author} {\bibfnamefont {T.}~\bibnamefont {Taniguchi}}, \bibinfo {author} {\bibfnamefont {J.}~\bibnamefont {Hone}}, \bibinfo {author} {\bibfnamefont {J.}~\bibnamefont {Shan}},\ and\ \bibinfo {author} {\bibfnamefont {K.~F.}\ \bibnamefont {Mak}},\ }\href {https://doi.org/10.1038/s41586-020-2868-6} {\bibfield  {journal} {\bibinfo  {journal} {Nature}\ }\textbf {\bibinfo {volume} {587}},\ \bibinfo {pages} {214} (\bibinfo {year} {2020})}\BibitemShut {NoStop}%
\bibitem [{\citenamefont {Guo}\ \emph {et~al.}(2024)\citenamefont {Guo}, \citenamefont {Pack}, \citenamefont {Swann}, \citenamefont {Holtzman}, \citenamefont {Cothrine}, \citenamefont {Watanabe}, \citenamefont {Taniguchi}, \citenamefont {Mandrus}, \citenamefont {Barmak}, \citenamefont {Hone}, \citenamefont {Millis}, \citenamefont {Pasupathy},\ and\ \citenamefont {Dean}}]{Guo2024Superconductivity}%
  \BibitemOpen
  \bibfield  {author} {\bibinfo {author} {\bibfnamefont {Y.}~\bibnamefont {Guo}}, \bibinfo {author} {\bibfnamefont {J.}~\bibnamefont {Pack}}, \bibinfo {author} {\bibfnamefont {J.}~\bibnamefont {Swann}}, \bibinfo {author} {\bibfnamefont {L.}~\bibnamefont {Holtzman}}, \bibinfo {author} {\bibfnamefont {M.}~\bibnamefont {Cothrine}}, \bibinfo {author} {\bibfnamefont {K.}~\bibnamefont {Watanabe}}, \bibinfo {author} {\bibfnamefont {T.}~\bibnamefont {Taniguchi}}, \bibinfo {author} {\bibfnamefont {D.}~\bibnamefont {Mandrus}}, \bibinfo {author} {\bibfnamefont {K.}~\bibnamefont {Barmak}}, \bibinfo {author} {\bibfnamefont {J.}~\bibnamefont {Hone}}, \bibinfo {author} {\bibfnamefont {A.~J.}\ \bibnamefont {Millis}}, \bibinfo {author} {\bibfnamefont {A.~N.}\ \bibnamefont {Pasupathy}},\ and\ \bibinfo {author} {\bibfnamefont {C.~R.}\ \bibnamefont {Dean}},\ }\href {https://doi.org/10.1038/s41586-024-08381-1} {\bibfield  {journal} {\bibinfo  {journal} {Nature}\ }\textbf {\bibinfo {volume} {626}},\ \bibinfo {pages} {754}
  (\bibinfo {year} {2024})}\BibitemShut {NoStop}%
\bibitem [{\citenamefont {Park}\ \emph {et~al.}(2023)\citenamefont {Park}, \citenamefont {Cai}, \citenamefont {Anderson}, \citenamefont {Zhang}, \citenamefont {Zhu}, \citenamefont {Liu}, \citenamefont {Wang}, \citenamefont {Holtzmann}, \citenamefont {Hu}, \citenamefont {Liu} \emph {et~al.}}]{park2023observation}%
  \BibitemOpen
  \bibfield  {author} {\bibinfo {author} {\bibfnamefont {H.}~\bibnamefont {Park}}, \bibinfo {author} {\bibfnamefont {J.}~\bibnamefont {Cai}}, \bibinfo {author} {\bibfnamefont {E.}~\bibnamefont {Anderson}}, \bibinfo {author} {\bibfnamefont {Y.}~\bibnamefont {Zhang}}, \bibinfo {author} {\bibfnamefont {J.}~\bibnamefont {Zhu}}, \bibinfo {author} {\bibfnamefont {X.}~\bibnamefont {Liu}}, \bibinfo {author} {\bibfnamefont {C.}~\bibnamefont {Wang}}, \bibinfo {author} {\bibfnamefont {W.}~\bibnamefont {Holtzmann}}, \bibinfo {author} {\bibfnamefont {C.}~\bibnamefont {Hu}}, \bibinfo {author} {\bibfnamefont {Z.}~\bibnamefont {Liu}}, \emph {et~al.},\ }\href {https://www.nature.com/articles/s41586-023-06536-0} {\bibfield  {journal} {\bibinfo  {journal} {Nature}\ }\textbf {\bibinfo {volume} {622}},\ \bibinfo {pages} {74} (\bibinfo {year} {2023})}\BibitemShut {NoStop}%
\bibitem [{\citenamefont {Li}\ \emph {et~al.}(2024)\citenamefont {Li}, \citenamefont {Dong}, \citenamefont {Ledwith},\ and\ \citenamefont {Khalaf}}]{li2024constraints}%
  \BibitemOpen
  \bibfield  {author} {\bibinfo {author} {\bibfnamefont {Q.}~\bibnamefont {Li}}, \bibinfo {author} {\bibfnamefont {J.}~\bibnamefont {Dong}}, \bibinfo {author} {\bibfnamefont {P.~J.}\ \bibnamefont {Ledwith}},\ and\ \bibinfo {author} {\bibfnamefont {E.}~\bibnamefont {Khalaf}},\ }\href {https://arxiv.org/abs/2407.02561} {\bibfield  {journal} {\bibinfo  {journal} {arXiv preprint arXiv:2407.02561}\ } (\bibinfo {year} {2024})}\BibitemShut {NoStop}%
\bibitem [{\citenamefont {Cole}\ and\ \citenamefont {Vanderbilt}(2024)}]{cole2024reduced}%
  \BibitemOpen
  \bibfield  {author} {\bibinfo {author} {\bibfnamefont {T.}~\bibnamefont {Cole}}\ and\ \bibinfo {author} {\bibfnamefont {D.}~\bibnamefont {Vanderbilt}},\ }\href {https://arxiv.org/abs/2412.17084} {\bibfield  {journal} {\bibinfo  {journal} {arXiv preprint arXiv:2412.17084}\ } (\bibinfo {year} {2024})}\BibitemShut {NoStop}%
\bibitem [{\citenamefont {Monsen}\ and\ \citenamefont {Claassen}(2024)}]{Monsen2024Supercell}%
  \BibitemOpen
  \bibfield  {author} {\bibinfo {author} {\bibfnamefont {B.}~\bibnamefont {Monsen}}\ and\ \bibinfo {author} {\bibfnamefont {M.}~\bibnamefont {Claassen}},\ }\href {https://arxiv.org/abs/2412.17190} {\bibfield  {journal} {\bibinfo  {journal} {arXiv preprint arXiv:2412.17190}\ } (\bibinfo {year} {2024})}\BibitemShut {NoStop}%
\bibitem [{\citenamefont {Girvin}\ \emph {et~al.}(1986)\citenamefont {Girvin}, \citenamefont {MacDonald},\ and\ \citenamefont {Platzman}}]{Girvin1987}%
  \BibitemOpen
  \bibfield  {author} {\bibinfo {author} {\bibfnamefont {S.~M.}\ \bibnamefont {Girvin}}, \bibinfo {author} {\bibfnamefont {A.~H.}\ \bibnamefont {MacDonald}},\ and\ \bibinfo {author} {\bibfnamefont {P.~M.}\ \bibnamefont {Platzman}},\ }\href {https://doi.org/10.1103/PhysRevB.33.2481} {\bibfield  {journal} {\bibinfo  {journal} {Phys. Rev. B}\ }\textbf {\bibinfo {volume} {33}},\ \bibinfo {pages} {2481} (\bibinfo {year} {1986})}\BibitemShut {NoStop}%
\bibitem [{\citenamefont {Haldane}(2011)}]{Haldane2011}%
  \BibitemOpen
  \bibfield  {author} {\bibinfo {author} {\bibfnamefont {F.~D.~M.}\ \bibnamefont {Haldane}},\ }\href {https://doi.org/10.1103/PhysRevLett.107.116801} {\bibfield  {journal} {\bibinfo  {journal} {Phys. Rev. Lett.}\ }\textbf {\bibinfo {volume} {107}},\ \bibinfo {pages} {116801} (\bibinfo {year} {2011})}\BibitemShut {NoStop}%
\bibitem [{\citenamefont {Halperin}\ and\ \citenamefont {Jain}(2020)}]{HalperinJain2020}%
  \BibitemOpen
  \bibinfo {editor} {\bibfnamefont {B.~I.}\ \bibnamefont {Halperin}}\ and\ \bibinfo {editor} {\bibfnamefont {J.~K.}\ \bibnamefont {Jain}},\ eds.,\ \href {https://doi.org/10.1142/11751} {\emph {\bibinfo {title} {Fractional Quantum Hall Effects: New Developments}}}\ (\bibinfo  {publisher} {World Scientific},\ \bibinfo {address} {Singapore},\ \bibinfo {year} {2020})\BibitemShut {NoStop}%
\bibitem [{\citenamefont {Roy}(2014)}]{Roy2014}%
  \BibitemOpen
  \bibfield  {author} {\bibinfo {author} {\bibfnamefont {R.}~\bibnamefont {Roy}},\ }\href {https://doi.org/10.1103/PhysRevB.90.165139} {\bibfield  {journal} {\bibinfo  {journal} {Phys. Rev. B}\ }\textbf {\bibinfo {volume} {90}},\ \bibinfo {pages} {165139} (\bibinfo {year} {2014})}\BibitemShut {NoStop}%
\bibitem [{\citenamefont {Ledwith}\ \emph {et~al.}(2021)\citenamefont {Ledwith}, \citenamefont {Khalaf},\ and\ \citenamefont {Vishwanath}}]{LEDWITH2021168646}%
  \BibitemOpen
  \bibfield  {author} {\bibinfo {author} {\bibfnamefont {P.~J.}\ \bibnamefont {Ledwith}}, \bibinfo {author} {\bibfnamefont {E.}~\bibnamefont {Khalaf}},\ and\ \bibinfo {author} {\bibfnamefont {A.}~\bibnamefont {Vishwanath}},\ }\href {https://doi.org/https://doi.org/10.1016/j.aop.2021.168646} {\bibfield  {journal} {\bibinfo  {journal} {Annals of Physics}\ }\textbf {\bibinfo {volume} {435}},\ \bibinfo {pages} {168646} (\bibinfo {year} {2021})},\ \bibinfo {note} {special issue on Philip W. Anderson}\BibitemShut {NoStop}%
\bibitem [{\citenamefont {Davidson}(2012)}]{davidson1976reduced}%
  \BibitemOpen
  \bibfield  {author} {\bibinfo {author} {\bibfnamefont {E.}~\bibnamefont {Davidson}},\ }\href {https://books.google.com/books?id=KX3qOxmTR9AC} {\emph {\bibinfo {title} {Reduced Density Matrices in Quantum Chemistry}}},\ Theoretical chemistry\ (\bibinfo  {publisher} {Academic Press},\ \bibinfo {year} {2012})\BibitemShut {NoStop}%
\bibitem [{\citenamefont {Yu}\ \emph {et~al.}(2024)\citenamefont {Yu}, \citenamefont {Bernevig}, \citenamefont {Queiroz}, \citenamefont {Rossi}, \citenamefont {T{\"o}rm{\"a}},\ and\ \citenamefont {Yang}}]{yu2024quantum}%
  \BibitemOpen
  \bibfield  {author} {\bibinfo {author} {\bibfnamefont {J.}~\bibnamefont {Yu}}, \bibinfo {author} {\bibfnamefont {B.~A.}\ \bibnamefont {Bernevig}}, \bibinfo {author} {\bibfnamefont {R.}~\bibnamefont {Queiroz}}, \bibinfo {author} {\bibfnamefont {E.}~\bibnamefont {Rossi}}, \bibinfo {author} {\bibfnamefont {P.}~\bibnamefont {T{\"o}rm{\"a}}},\ and\ \bibinfo {author} {\bibfnamefont {B.-J.}\ \bibnamefont {Yang}},\ }\href {https://arxiv.org/abs/2501.00098} {\bibfield  {journal} {\bibinfo  {journal} {arXiv preprint arXiv:2501.00098}\ } (\bibinfo {year} {2024})}\BibitemShut {NoStop}%
\bibitem [{\citenamefont {Mera}\ and\ \citenamefont {Ozawa}(2022)}]{Mera2022}%
  \BibitemOpen
  \bibfield  {author} {\bibinfo {author} {\bibfnamefont {B.}~\bibnamefont {Mera}}\ and\ \bibinfo {author} {\bibfnamefont {T.}~\bibnamefont {Ozawa}},\ }\href {https://doi.org/10.1103/PhysRevB.106.245134} {\bibfield  {journal} {\bibinfo  {journal} {Phys. Rev. B}\ }\textbf {\bibinfo {volume} {106}},\ \bibinfo {pages} {245134} (\bibinfo {year} {2022})}\BibitemShut {NoStop}%
\bibitem [{\citenamefont {Wieder}\ \emph {et~al.}(2016)\citenamefont {Wieder}, \citenamefont {Kim}, \citenamefont {Rappe},\ and\ \citenamefont {Kane}}]{wieder2016double}%
  \BibitemOpen
  \bibfield  {author} {\bibinfo {author} {\bibfnamefont {B.~J.}\ \bibnamefont {Wieder}}, \bibinfo {author} {\bibfnamefont {Y.}~\bibnamefont {Kim}}, \bibinfo {author} {\bibfnamefont {A.~M.}\ \bibnamefont {Rappe}},\ and\ \bibinfo {author} {\bibfnamefont {C.~L.}\ \bibnamefont {Kane}},\ }\href {https://doi.org/10.1103/PhysRevLett.116.186402} {\bibfield  {journal} {\bibinfo  {journal} {Phys. Rev. Lett.}\ }\textbf {\bibinfo {volume} {116}},\ \bibinfo {pages} {186402} (\bibinfo {year} {2016})}\BibitemShut {NoStop}%
\bibitem [{\citenamefont {Bradlyn}\ \emph {et~al.}(2016)\citenamefont {Bradlyn}, \citenamefont {Cano}, \citenamefont {Wang}, \citenamefont {Vergniory}, \citenamefont {Felser}, \citenamefont {Cava},\ and\ \citenamefont {Bernevig}}]{bradlyn2016beyond}%
  \BibitemOpen
  \bibfield  {author} {\bibinfo {author} {\bibfnamefont {B.}~\bibnamefont {Bradlyn}}, \bibinfo {author} {\bibfnamefont {J.}~\bibnamefont {Cano}}, \bibinfo {author} {\bibfnamefont {Z.}~\bibnamefont {Wang}}, \bibinfo {author} {\bibfnamefont {M.~G.}\ \bibnamefont {Vergniory}}, \bibinfo {author} {\bibfnamefont {C.}~\bibnamefont {Felser}}, \bibinfo {author} {\bibfnamefont {R.~J.}\ \bibnamefont {Cava}},\ and\ \bibinfo {author} {\bibfnamefont {B.~A.}\ \bibnamefont {Bernevig}},\ }\href {https://doi.org/10.1126/science.aaf5037} {\bibfield  {journal} {\bibinfo  {journal} {Science}\ }\textbf {\bibinfo {volume} {353}},\ \bibinfo {pages} {aaf5037} (\bibinfo {year} {2016})}\BibitemShut {NoStop}%
\bibitem [{\citenamefont {Tan}\ and\ \citenamefont {Devakul}(2024)}]{Tan2024}%
  \BibitemOpen
  \bibfield  {author} {\bibinfo {author} {\bibfnamefont {T.}~\bibnamefont {Tan}}\ and\ \bibinfo {author} {\bibfnamefont {T.}~\bibnamefont {Devakul}},\ }\href {https://doi.org/10.1103/PhysRevX.14.041040} {\bibfield  {journal} {\bibinfo  {journal} {Phys. Rev. X}\ }\textbf {\bibinfo {volume} {14}},\ \bibinfo {pages} {041040} (\bibinfo {year} {2024})}\BibitemShut {NoStop}%
\bibitem [{\citenamefont {T{\"o}rm{\"a}}\ \emph {et~al.}(2022)\citenamefont {T{\"o}rm{\"a}}, \citenamefont {Peotta},\ and\ \citenamefont {Bernevig}}]{torma2022}%
  \BibitemOpen
  \bibfield  {author} {\bibinfo {author} {\bibfnamefont {P.}~\bibnamefont {T{\"o}rm{\"a}}}, \bibinfo {author} {\bibfnamefont {S.}~\bibnamefont {Peotta}},\ and\ \bibinfo {author} {\bibfnamefont {B.~A.}\ \bibnamefont {Bernevig}},\ }\href {https://doi.org/10.1038/s42254-022-00466-y} {\bibfield  {journal} {\bibinfo  {journal} {Nat Rev Phys}\ }\textbf {\bibinfo {volume} {4}},\ \bibinfo {pages} {528} (\bibinfo {year} {2022})}\BibitemShut {NoStop}%
\bibitem [{\citenamefont {Yu}\ \emph {et~al.}(2023)\citenamefont {Yu}, \citenamefont {Xie}, \citenamefont {Bernevig},\ and\ \citenamefont {Sarma}}]{Yu2023}%
  \BibitemOpen
  \bibfield  {author} {\bibinfo {author} {\bibfnamefont {J.}~\bibnamefont {Yu}}, \bibinfo {author} {\bibfnamefont {M.}~\bibnamefont {Xie}}, \bibinfo {author} {\bibfnamefont {B.~A.}\ \bibnamefont {Bernevig}},\ and\ \bibinfo {author} {\bibfnamefont {S.~D.}\ \bibnamefont {Sarma}},\ }\href {https://doi.org/10.1103/PhysRevB.108.035129} {\bibfield  {journal} {\bibinfo  {journal} {Phys. Rev. B}\ }\textbf {\bibinfo {volume} {108}},\ \bibinfo {pages} {035129} (\bibinfo {year} {2023})}\BibitemShut {NoStop}%
\bibitem [{\citenamefont {Herzog-Arbeitman}\ \emph {et~al.}(2024)\citenamefont {Herzog-Arbeitman}, \citenamefont {Yu}, \citenamefont {C\u{a}lug\u{a}ru}, \citenamefont {Hu}, \citenamefont {Regnault}, \citenamefont {Liu}, \citenamefont {Sarma}, \citenamefont {Vafek}, \citenamefont {Coleman}, \citenamefont {Tsvelik}, \citenamefont {da~Song},\ and\ \citenamefont {Bernevig}}]{Herzog-Arbeitman2024}%
  \BibitemOpen
  \bibfield  {author} {\bibinfo {author} {\bibfnamefont {J.}~\bibnamefont {Herzog-Arbeitman}}, \bibinfo {author} {\bibfnamefont {J.}~\bibnamefont {Yu}}, \bibinfo {author} {\bibfnamefont {D.}~\bibnamefont {C\u{a}lug\u{a}ru}}, \bibinfo {author} {\bibfnamefont {H.}~\bibnamefont {Hu}}, \bibinfo {author} {\bibfnamefont {N.}~\bibnamefont {Regnault}}, \bibinfo {author} {\bibfnamefont {C.}~\bibnamefont {Liu}}, \bibinfo {author} {\bibfnamefont {S.~D.}\ \bibnamefont {Sarma}}, \bibinfo {author} {\bibfnamefont {O.}~\bibnamefont {Vafek}}, \bibinfo {author} {\bibfnamefont {P.}~\bibnamefont {Coleman}}, \bibinfo {author} {\bibfnamefont {A.}~\bibnamefont {Tsvelik}}, \bibinfo {author} {\bibfnamefont {Z.}~\bibnamefont {da~Song}},\ and\ \bibinfo {author} {\bibfnamefont {B.~A.}\ \bibnamefont {Bernevig}},\ }\href {https://arxiv.org/abs/2404.07253} {\bibfield  {journal} {\bibinfo  {journal} {arXiv preprint arXiv:2404.07253}\ } (\bibinfo {year} {2024})}\BibitemShut {NoStop}%
\bibitem [{\citenamefont {Ledwith}\ \emph {et~al.}(2024)\citenamefont {Ledwith}, \citenamefont {Dong}, \citenamefont {Vishwanath},\ and\ \citenamefont {Khalaf}}]{ledwith2024nonlocal}%
  \BibitemOpen
  \bibfield  {author} {\bibinfo {author} {\bibfnamefont {P.~J.}\ \bibnamefont {Ledwith}}, \bibinfo {author} {\bibfnamefont {J.}~\bibnamefont {Dong}}, \bibinfo {author} {\bibfnamefont {A.}~\bibnamefont {Vishwanath}},\ and\ \bibinfo {author} {\bibfnamefont {E.}~\bibnamefont {Khalaf}},\ }\href {https://arxiv.org/abs/2408.16761} {\bibfield  {journal} {\bibinfo  {journal} {arxiv:2408.16761}\ } (\bibinfo {year} {2024})}\BibitemShut {NoStop}%
\bibitem [{\citenamefont {Bernevig}\ \emph {et~al.}(2021{\natexlab{e}})\citenamefont {Bernevig}, \citenamefont {Lian}, \citenamefont {Cowsik}, \citenamefont {Xie}, \citenamefont {Regnault},\ and\ \citenamefont {Song}}]{TBG5}%
  \BibitemOpen
  \bibfield  {author} {\bibinfo {author} {\bibfnamefont {B.~A.}\ \bibnamefont {Bernevig}}, \bibinfo {author} {\bibfnamefont {B.}~\bibnamefont {Lian}}, \bibinfo {author} {\bibfnamefont {A.}~\bibnamefont {Cowsik}}, \bibinfo {author} {\bibfnamefont {F.}~\bibnamefont {Xie}}, \bibinfo {author} {\bibfnamefont {N.}~\bibnamefont {Regnault}},\ and\ \bibinfo {author} {\bibfnamefont {Z.-D.}\ \bibnamefont {Song}},\ }\href {https://doi.org/10.1103/PhysRevB.103.205415} {\bibfield  {journal} {\bibinfo  {journal} {Phys. Rev. B}\ }\textbf {\bibinfo {volume} {103}},\ \bibinfo {pages} {205415} (\bibinfo {year} {2021}{\natexlab{e}})}\BibitemShut {NoStop}%
\end{thebibliography}%

\begin{appendix}

\onecolumngrid
\newpage
\makeatletter

\setcounter{figure}{0}
\setcounter{section}{0}
\setcounter{equation}{0}

\renewcommand{\thefigure}{S\@arabic\c@figure}
\renewcommand{\thesection}{S-\@Roman\c@section}
\makeatother

\appendix

\section{Charge density}
The charge density operator is written in terms of electron field operators $\Psi_r$ in the continuum
\begin{equation}
    \hat{\rho}(r) =  \Psi^\dag_r \Psi^\phdag_r
\end{equation}
where $r$ is a continuous variable.
In the tight-binding approximation, we discretize this quantity in terms of microscopic orbitals
\begin{equation}
    \Psi^\phdag_r = \sum\limits_{\alpha, R} \phi_\alpha(r - R) \;d^\phdag_{\alpha, R}
\end{equation}
where $\phi_\alpha(r)$ are rapidly decaying functions of $r$ and $R$ is the span of lattice vectors. The density then becomes
\begin{equation}
    \hat{\rho}(r) = \sum\limits_{\alpha, R} \sum\limits_{\beta, R^\prime} \phi^*_\alpha(r - R) \phi_\beta(r - R^\prime) \;d^\dag_{\alpha, R} d^\phdag_{\beta, R^\prime}
\end{equation}
which can be integrated to get the total density
\begin{equation}
    \hat{\rho} = \int dr \hat{\rho}(r) = \sum\limits_{\alpha, R} \sum\limits_{\beta, R^\prime} \left(\int dr \phi^*_\alpha(r - R) \phi_\beta(r - R^\prime)\right) \;d^\dag_{\alpha, R} d^\phdag_{\beta, R^\prime}
\end{equation}
where the spatial integral gives $\delta_{R,R^\prime} \delta_{\alpha\beta}$ and we get
\begin{equation}
    \hat{\rho} = \sum\limits_{\alpha, R} d^\dag_{\alpha, R} d^\phdag_{\alpha, R}.\label{eq:total-rho}
\end{equation}
From this expression of total charge density, we now define the total density inside inside one unit cell as
\begin{equation}
    \hat{\rho}_R \equiv \sum\limits_{ \alpha} d^\dag_{\alpha, R} d^\phdag_{\alpha, R}. \label{eq:def-proj-rho-R}
\end{equation}
We pause here to comment on two subtle aspects of this definition. First, it depends on the choice of unit cell. While this ambiguity does not affect the total charge density in Eq.~\eqref{eq:total-rho}, different choices of how the total charge is partitioned among unit cells can give different local charge distributions. This point will be explored in detail in an upcoming work.
The second subtlety is the independence from orbital embedding. While it is clear from Eq.~\eqref{eq:def-proj-rho-R}, it is instructive to confirm embedding independence in the projected density operator used in the main text.
To illustrate this, we keep track of the location of the orbital $\tau_\alpha$, and transform to momentum space with 
\begin{equation}
    d_{\alpha, R} = \dfrac{1}{\sqrt{N}} \sum\limits_{k \in {\rm BZ}} e^{ i k\cdot (R+\tau_\alpha)} d_{\alpha, k,}
\end{equation}
that leads to
\begin{equation}
    \hat{\rho}_R \equiv \dfrac{1}{N}\sum\limits_{k,k^\prime} \sum\limits_{ \alpha} e^{ i (k-k^\prime)\cdot(R+\tau_\alpha)} d^\dag_{\alpha, k} d^\phdag_{\alpha, k^\prime}
\end{equation}
from which we can transform to band basis using the transformation
\begin{equation}
    d^\phdag_{\alpha, k} = \sum\limits_m u_{m,\alpha}(k) c^\phdag_{m, k}
\end{equation}
to get
\begin{equation}
    \hat{\rho}_R \equiv \dfrac{1}{N}\sum\limits_{k,k^\prime} \sum\limits_{m,n}
    \left( \sum\limits_{ \alpha} u^*_{m,\alpha}(k) e^{ i (k-k^\prime)\cdot(R+\tau_\alpha)} u_{n,\alpha}(k^\prime) \right) c^\dag_{m,k} c^\phdag_{n, k^\prime}
\end{equation}
where we can define the inner product
\begin{equation}
    \sum\limits_{ \alpha} u^*_{m,\alpha}(k) e^{ i (k-k^\prime)\cdot(R+\tau_\alpha)} u_{n,\alpha} \equiv e^{ i (k-k^\prime)\cdot R} \langle u_{m, k} | e^{ i (k-k^\prime)\cdot \hat{r}} | u_{n, k^\prime} \rangle \label{eq:inner-prod}
\end{equation}
and arrive at 
\begin{equation}
    \hat{\rho}_R \equiv \dfrac{1}{N}\sum\limits_{k,k^\prime} \sum\limits_{m,n} e^{ i (k-k^\prime)\cdot R} \langle u_{m, k} | e^{ i (k-k^\prime)\cdot \hat{r}} | u_{n, k^\prime} \rangle
     c^\dag_{m,k} c^\phdag_{n,k^\prime}.
\end{equation}

The projected density operator is then obtained by restricting the summation to the active band, say $n$, 
\begin{equation}
    \hat{\rho}_R \rightarrow \hat{\bar{\rho}}_R = \dfrac{1}{N}\sum\limits_{k,k^\prime \in {\rm BZ}} e^{ i (k-k^\prime)\cdot R} \langle u_{n,k} | e^{ i (k-k^\prime)\cdot \hat{r}} | u_{n,k^\prime} \rangle c^\dag_{n,k} c^\phdag_{n,k^\prime} = \sum\limits_{k,k^\prime \in {\rm BZ}} e^{ i (k-k^\prime)\cdot R} c^\dag_{n, k} [\Lambda]_{k,k^\prime} c^\phdag_{n,k^\prime}.
\end{equation}
The effect of orbital embedding can now be understood from the matrix elements. If $k^\prime = k+ G$ where $G$ is a reciprocal lattice vector, we know that embedding enforces 
\begin{equation}
    |u_{k+G} \rangle = V_G | u_k\rangle
\end{equation}
where $V_G = {\rm diag}.[\cdots e^{ i G\cdot \tau_\alpha}\cdots ]$ is the embedding matrix. This additional factor is exactly canceled by the exponential in Eq.~\eqref{eq:inner-prod} and the net matrix element is independent of the choice of $\tau_\alpha$.
In what follows next, we assume trivial embedding to simplify calculations and set all $\tau_\alpha$ to zero.

The unit cell label $R$ can be absorbed into a unitary matrix that rotates the $c_{n,k}$ operators. It does not change the eigenvalues of the projected density operator but gives a phase to the eigenvector. As a result we can write the projected density as
\begin{equation}
    \hat{\bar{\rho}}_R = \sum\limits_{a=1}^N \lambda_a c^\dag_{a,R} c^\phdag_{a, R}
\end{equation}
where the modes are defined as
\begin{equation}
    c^\phdag_{a, R} = \sum\limits_{k \in {\rm BZ}} e^{ i k\cdot R} s_{a, k} c^\phdag_{n, k}.
\end{equation}
The envelope function $s_{a,k}$ and the mode values $\lambda_a$ are obtained from eigen decomposition of $\Lambda_{k,k^\prime}$.

\end{appendix}

\end{document}